\documentclass[12pt]{iopart}
\usepackage{iopams}
\usepackage{setstack}
\usepackage[colorlinks=true,citecolor=blue,linkcolor=magenta]{hyperref}
\usepackage{graphicx}
\usepackage{bm}
\usepackage{dsfont}
\usepackage{bbm}
\newcommand{\bra}[1]{\langle #1|}
\newcommand{\ket}[1]{|#1\rangle}
\newcommand{\ex}[1]{\langle #1 \rangle}

\begin{document}
\title{Changing optical band structure with single photons}

\author{Andreas Albrecht, Tommaso Caneva, and Darrick E Chang}
\address{ICFO - Institut de Ci\`encies Fot\`oniques, The Barcelona Institute of Science and Technology, 08860 Castelldefels (Barcelona), Spain.}

\begin{abstract}
Achieving strong interactions between individual photons enables a wide variety of exciting possibilities in quantum information science and many-body physics. Cold atoms interfaced with nanophotonic structures have emerged as a platform to realize novel forms of nonlinear interactions. In particular, when atoms are coupled to a photonic crystal waveguide (PCW), long-range atomic interactions can arise that are mediated by localized atom-photon bound states. 
We theoretically show that in such a system,  the absorption of a single photon can change the band structure for a subsequent photon. This occurs because the first photon affects the atoms in the chain in an alternating fashion, thus leading to an effective period doubling of the system and a new optical band structure for the composite atom-nanophotonic system.
 We demonstrate how this mechanism can be engineered to realize a single-photon switch, where the first incoming photon switches  the system from being highly transmissive to highly reflective, and analyze how signatures can be observed via non-classical correlations of the outgoing photon field. 
\end{abstract}

\maketitle

\section{Introduction}
Creating strong, controllable interactions between individual photons enables many opportunities\,\cite{chang14rev} ranging  from quantum information processing to the creation of strongly correlated quantum states of light\,\cite{hartmann06, greentree06, chang08, angelakis07, noh16}.   Atoms form ideal  nonlinear photonic elements to accomplish these tasks, and therefore have been used in diverse settings as cavity QED\,\cite{turchette95, kimble08, reiserer15} or Rydberg atomic gases\,\cite{pritchard10, peyronel12, pritchard13}.  A promising new approach is the field of waveguide QED\,\cite{chang07, loo13}, made possible by experiments to interface atoms with propagating modes of nanophotonic systems including nanofibers\,\cite{vetsch10} and photonic crystal waveguides (PCWs)\,\cite{goban14, goban15, javadi15, lodahl15}.
While these platforms were originally developed to improve upon figures of merit compared to their macroscopic counterparts, more interestingly they enable unprecedented possibilities for engineering interactions with no obvious prior analogue\,\cite{petersen14, douglas15}.

Here, we identify and analyze a remarkable example, where a single propagating photon effectively changes the photonic band structure for a second photon. This effect relies on the fact that a near-resonant photon is efficiently converted into an atomic excitation, and that atomic excitations can be made to interact strongly and coherently by aligning the atomic transition frequency with band gaps of an underlying PCW\,\cite{douglas15, john90, kurizki90}.
Specifically, an excited atom hybridizes with a localized photon to form an atom-photon bound state, the photonic component of which can couple to proximal atoms\,\cite{douglas15, hood16, liu16}.
This mechanism has been validated experimentally only recently for atoms along PCWs\,\cite{hood16} and transmon qubits\,\cite{liu16}. Here, we exploit that for an atomic lattice trapped with the PCW periodicity, atom-atom interactions naturally alternate in sign due to the Bloch structure of the underlying optical modes. Thus, the presence of one atomic excitation effectively creates a doubling of the atomic periodicity from the standpoint of a second propagating photon, resulting in a dramatic change of its dispersion relation.  This should be contrasted with the qualitatively very different physics that arises for the case of spatially smooth atomic interactions, which has been separately analyzed before for PCWs\,\cite{douglas15mol, shahmoon14} and also lies at the heart of optical nonlinearities involving Rydberg gases\,\cite{pritchard10, peyronel12, pritchard13}.

We focus specifically on a regime in which a periodicity doubling leads to a so-called atomic-mirror configuration \,\cite{chang12, corzo16, sorensen16}. This creates a situation in which the absorption of a single photon makes the medium highly reflecting for a subsequent one. 
In Section\,\ref{sec_config} we present the detailed setup of our proposal and outline how a single photon influences the propagation of subsequent photons. In Section\,\ref{sec_lshift}, we discuss a particular scheme, in which the atom-atom interactions arising from a PCW bandgap can be utilized to create a dispersive interaction, where an excited atom shifts the resonance frequencies of proximal atoms in an alternating fashion. Subsequently, Section\,\ref{sec_correl} discusses the effect of these atomic interactions on light propagation. Specifically, we show how the high conditional reflectance produced by the absorption of a single photon manifests itself in the form of an anti-bunched transmitted field, given a weak classical input state. An application of the mechanism as a single-photon switch is discussed in Section\,\ref{sec_condphoton},  in which a single-photon wavepacket is first mapped into the system to regulate the propagation of subsequent photons. This is followed by a discussion on how decoherence impacts the photon switch in  Section\,\ref{sec_dephasing}. 

\section{Dynamic atomic mirror configuration}\label{sec_config}
\begin{figure}[tbh]
\begin{centering}
\includegraphics[scale=0.3]{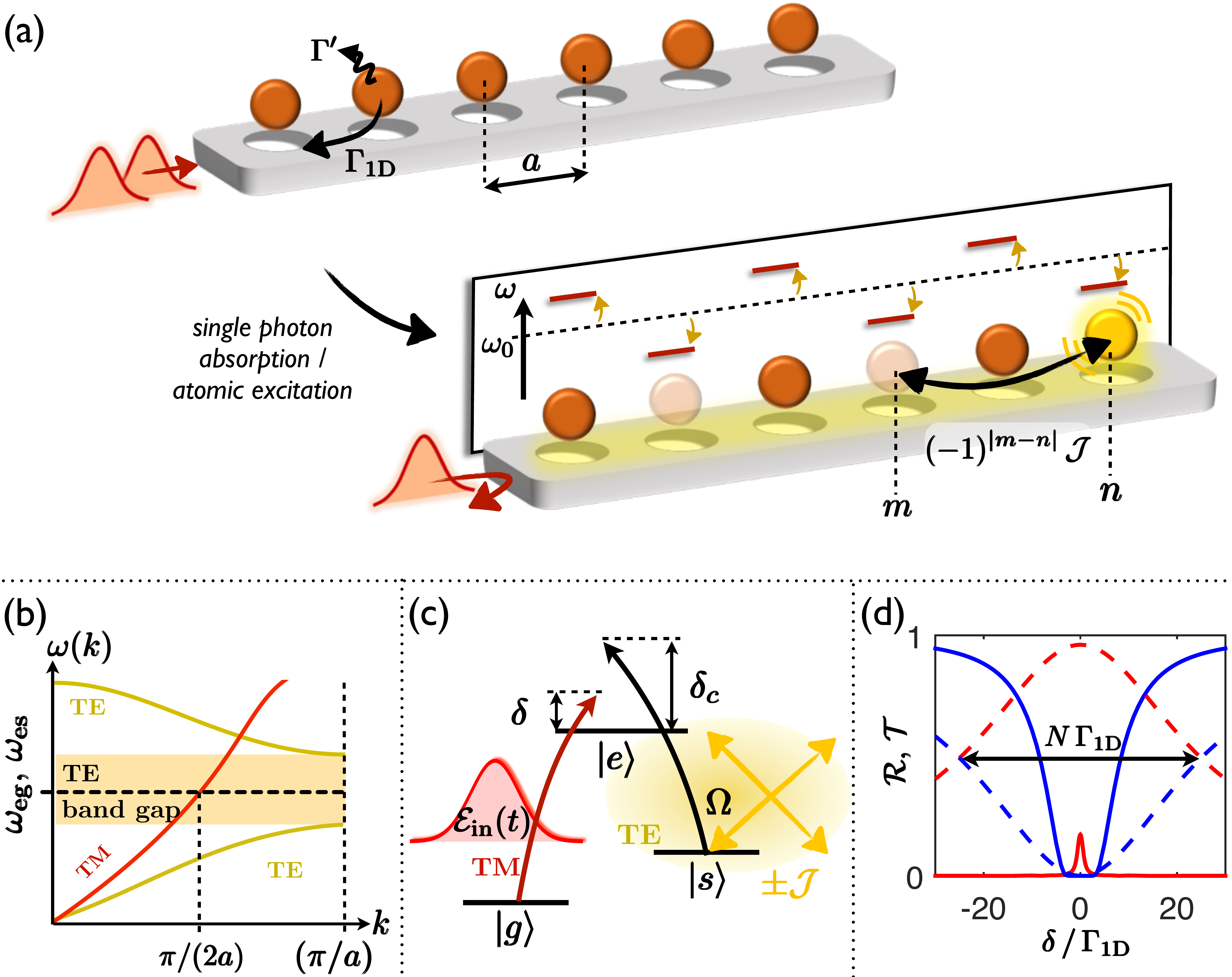}
\caption{\label{b_setup} (a) Array of atoms along a PCW, trapped with the structure period $a$.  Lower: An excitation (single photon absorption, yellow atom) induces strong, long-range and alternating atom-atom interactions $\mathcal{J}$ mediated by a photonic bound state within the PCW band gap (yellow). In certain regimes, this interaction effectively produces an alternating shift of the atomic transition frequency $\omega$ from its bare value $\omega_0$ as indicated by the vertical scale. (b) Schematic band structure of the TE and TM mode. The horizontal dashed line marks the atomic transition frequencies, here without loss of generality drawn equal. (c) Atomic three-level structure and couplings (see main text for definitions). (d)  Reflectance (red) and transmittance (blue) for an input field  detuning $\delta$ from a two-level transition and $\Gamma_{\rm 1D}=\Gamma'$. Solid lines correspond to $N=100$, $k\,a=\pi/2$, dashed lines to a sublattice of every second atom $N=50$, $k\,a=\pi$.   }
\end{centering}
\end{figure}

We consider an array of $N$ atoms trapped along a PCW with the photonic crystal periodicity $a$ (Fig.\,\ref{b_setup}\,(a)).
We assume that the $\ket{g}$ to $\ket{e}$ transition of an atomic three-level system, with frequency $\omega_{\rm eg}$,  couples to a guided mode of given polarization (e.g., transverse magnetic (TM), Fig.\,\ref{b_setup}\,(b),\,(c)). In line with experiments, we assume a single atom can emit into the waveguide at a rate $\Gamma_{\rm 1D}$ significant in magnitude compared to its free-space rate $\Gamma'$\,\cite{ goban14, goban15}. 
We probe the $g$-$e$ transition by a (weak coherent) input field of the form $\mathcal{E}_{\rm in}(z,t)=\tilde{\mathcal{E}}_{\rm in}(t)e^{i(kz-\delta t)}$ through the waveguide, where $k$ denotes the wavevector, $\delta$ the detuning from atomic resonance and $\tilde{\mathcal{E}}_{\rm in}(t)$ accounts for the temporal pulse shape. Its propagation can be described in a  spin-model description\,\cite{chang12, caneva15, lekien05, dzsotjan10, tudela11}. In that framework the atomic dynamics are governed by the Hamiltonian $\mathcal{H}_{\rm TM}=\mathcal{H}_{\rm 0}+\mathcal{H}_{\rm wg}$. Here, $\mathcal{H}_{0}$ describes the atom coupling to the input field and the free-space decay and takes the form 
\begin{equation} \mathcal{H}_{0}=-i\frac{\Gamma'}{2}\sum_m \sigma_{\rm ee}^m-\sum_m \left[\tilde{\mathcal{E}}_{\rm in}(t)\,e^{ikz_m}\,e^{-i\delta\,t}\sigma_{\rm eg}^m+\mathrm{h.c.}\right] , \end{equation}
with $z_m=m\,a$ the position of atom $m$ and the atomic operators defined as $\sigma_{ij}=\ket{i}\bra{j}$.
The spin Hamiltonian $\mathcal{H}_{\rm wg}=-i\frac{\Gamma_{\rm 1D}}{2}\sum_{m,n} e^{i\,k |z_m-z_n|}\,\sigma_{\rm eg}^m\,\sigma_{\rm ge}^n$ accounts for the coupling of the atoms via the emission and re-absorption of photons in the TM mode. Such an effective spin Hamiltonian description for the atom-photon interaction is valid in the Markovian limit,  where time retardation over the typical atomic bandwidth frequency range and associated with the free field waveguide propagation is negligible, an assumption readily fulfilled for typical atomic chain sizes\,\cite{caneva15, shi15}. More general non-Markovian approaches and the validity range of the Markovian limit for a variety of atom-waveguide configurations have been studied in the literature\,\cite{shi15, guimond16, zheng13, ballestero13, pletyukhov12}. Once the atomic dynamics are determined from evolution under $\mathcal{H}_{\rm TM}$, the right (+)\,- and left (-)\,-going field can be recovered using the input-output relations\,\cite{chang12, caneva15, lalumiere13, xu15} $	\mathcal{E}_{\pm}(z,t) =\mathcal{E}_{\rm in}^{\pm}(z,t)+i\,\frac{\Gamma_{\rm 1D}}{2}\,\sum_{z_m}\sigma_{\rm ge}^m e^{\pm i\,k(z-z_m)} $, which relate the outgoing field in terms of the atomic coherences and input fields $\mathcal{E}_{\rm in}^{\pm}$ (here $\mathcal{E}_{\rm in}^+(z,t)\equiv\mathcal{E}_{\rm in}(z,t)$, while $\mathcal{E}_{\rm in}^{-}(z,t)=0$).

A second transition involving metastable state $\ket{s}$ and $\ket{e}$ is assumed to have a different polarization, and couples to the transverse electric (TE) mode of the waveguide. The transition frequency $\omega_{\rm es}$ is assumed to lie within a TE band gap (see Fig.\,\ref{b_setup}\,(b)).
Thus, $\ket{e}$ cannot spontaneously emit into the band due to the absence of resonant modes. Instead, the photon that the atom attempts to emit is exponentially confined to a length $L$ around the atomic position\,\cite{douglas15, john90}. This photon can be coherently exchanged with proximal atoms, leading to a long-range interaction of the form\,\cite{douglas15, kurizki90, hood16}   
	\begin{equation}\label{hambg} \mathcal{H}_{\rm bg}=\mathcal{J}\,\sum_{m\neq n}\cos(q\,z_m)\cos(q\,z_n)\,e^{-|z_m-z_n|/L}\,\sigma_{\rm es}^m\sigma_{\rm se}^n\,.  
	\end{equation}
Here, the cosine terms account for the standing-wave nature of the field at frequencies within a band gap\,\cite{goban15}, in direct analogy to a Fabry-Perot cavity.  Within the TE band gap the wavevector $q=\pi/a$, such that $\cos(qz_m)\cos(qz_n)=(-1)^{(m+n)}$ and the interaction is intrinsically of alternating sign. Whereas previous proposals neglected that sign alternation\,\cite{douglas15mol,shahmoon14, caneva15}, it forms the crucial mechanism for creating non-linear photon interactions in our setup. The alternating character also  makes our approach clearly distinct from setups based on Rydberg atoms\,\cite{pritchard10, peyronel12, pritchard13}, where the photon nonlinearities arise from spatially smooth atom-atom interactions.   We moreover assume that the longitudinal spatial extent is tuned to be nearly constant\,\cite{douglas15, hood16}, such that $\exp(-|z_m-z_n |/L)\simeq 1$. 
On one hand, this limit is readily achievable experimentally\,\cite{hood16}, but conceptually it also serves as a useful baseline, given that alternating interactions have never been explored from the standpoint of nonlinear optics.  In addition, we introduce a control field $\Omega$ from free-space on the band gap ($s$-$e$) transition that allows for modifying the atom-atom interaction character. It can be described by $\mathcal{H}_c=\delta_c \sum_m\sigma_{\rm ss}^m-\Omega\sum_m\left[\sigma_{\rm es}^m+\mathrm{h.c.}\right]$ with $\delta_c$ the control field detuning from the $s$-$e$ resonance. Thus the total Hamiltonian is $\mathcal{H}=\mathcal{H}_{\rm TM}+\mathcal{H}_c+\mathcal{H}_{\rm bg}$, which combined with the input-output operators allows for solving the probe field propagation dynamics.

Note that, although an alternating interaction like (\ref{hambg}) can also be achieved in a cavity (where $L\rightarrow\infty$)\,\cite{goldstein97}, there such a coupling would naturally be probed by the output field associated with the same mode that generated the interactions. In contrast, the PCW offers the opportunity to effectively combine cavity QED and waveguide QED. In particular, the bandgap interaction produced by the TE mode can be probed by propagating fields in a co-linear TM mode. As we will see, while the atoms would respond to propagating TM photons alone like a typical atomic ensemble, the strong TE-mediated interactions will endow the propagation with novel quantum nonlinearities.

We assume that the atoms are initialized in state $\ket{g}$, and we restrict the analysis to two excitations in the $\{\ket{e},\ket{s}\}$ manifold. This corresponds to weak input states in the TM mode where the probability of more than two photons being absorbed is negligible.  Spin-flip interactions (\ref{hambg}) require at least one excitation in each of the states $\ket{e}$ and $\ket{s}$. Since all atoms are initialized in $\ket{g}$, a single photon propagating in the TM mode remains unaffected by the band gap interaction.
To understand two-photon propagation, we examine the atomic two-excitation spectrum. In the regime $|\delta_c|\gtrsim |\Omega|, |\mathcal{J}|$, the states $\ket{e}$ and $\ket{s}$ are not effectively mixed together. The states $\ket{e_m e_n}$ and $\ket{s_m s_n}$ therefore remain eigenstates within lowest-order perturbation theory, but experience an energy shift as the far off-resonant control field weakly couples these states to states  $(\ket{s_m e_n}+\ket{e_m s_n})/\sqrt{2}$. Due to $\mathcal{H}_{\rm bg}$ the energy difference between these two states is not just $\delta_c$, but shifted additionally by $\mathcal{J}_{mn}=(-1)^{m+n}\,\mathcal{J}$. This results in ac-Stark energy shifts, which in section\,\ref{sec_lshift} are shown to be  $\Delta\omega_{\rm ee}=-2\Omega^2/(\delta_c+\mathcal{J}_{mn})$ and $\Delta\omega_{\rm ss}=2\Omega^2/(\delta_c-\mathcal{J}_{\rm mn})$  for $\ket{e_m e_n}$ and $\ket{s_m s_n}$, respectively.

Consequently, two-excitation energy levels pick up an alternation due to the underlying Bloch modes.  The transition energies can thus be divided into two sublattices of double periodicity\,(Fig.\,\ref{b_setup}\,(a)), separated in energy by $\Delta\nu\simeq4\Omega^2 \mathcal{J}/(\delta_c^2-\mathcal{J}^2)$.  We now discuss how this manifests itself in photon propagation.
 Suppose a first photon is initially sent into the medium. The photon has a high chance of creating an atomic excitation in $\ket{e}$ ($\ket{s}$), if its frequency is close to single-photon (two-photon) resonance $\delta\sim 0$ ($\Delta\equiv(\delta-\delta_c)\sim 0$). The propagation of a second photon through the medium is then governed by the frequencies of the doubly excited states $\ket{e_m e_n}$ (`e-branch') or $\ket{s_m s_n}$ (`s-branch'). The second photon thus sees a new band structure associated with this atomic ``refractive index'' pattern of doubled periodicity. In particular, if the splitting is resolved $\Delta\nu >\Gamma'$, the second photon would effectively either see a highly transparent medium if its frequency is similar to the bare atomic transition (as all resonances have been shifted), or see only every other atom if its frequency is aligned with one of the sublattice transitions.

Even though that concept is quite general, we here choose $k\,a=\pi/2$, in which case the period-doubled configuration $k\,(2a)=\pi$ is known to result in a dramatically different photon propagation.  Whereas  $k\,a=\pi/2$ represents a configuration that minimizes reflection\,\cite{douglas15mol, caneva15}, $k\,a=\pi$ corresponds to the atomic-mirror configuration of constructive reflection\,\cite{chang12, corzo16, sorensen16}.  In particular, for an array of $N/2$ two-level atoms (i.e. states $g$ and $e$ evolving under $\mathcal{H}_0+\mathcal{H}_{\rm wg}$) with this spacing, the reflectance spectrum is characterized by a Lorentzian\,\cite{chang12}
\begin{equation}\label{refl1}  \mathcal{R}(\Gamma_{\rm 1D}, \Gamma', \delta)=\frac{\left[(N/2)\,\Gamma_{\rm 1D}\right]^2}{\left[(N/2)\,\Gamma_{\rm 1D}+\Gamma'\right]^2+(2\delta)^2}\,.   \end{equation} 
Here the reflectance $\mathcal{R}$ is defined as the ratio of reflected to input intensity. For $\Gamma'\ll(N/2)\,\Gamma_{\rm 1D}$, the resonance width scales like $\Delta\omega=(N/2)\,\Gamma_{\rm 1D}$. This property arises from the superradiant collective response of the atoms\,\cite{chang12}, and is in fact independent of the filling fraction of trapped atoms. Thus, our choice of lattice constant $k\,a=\pi/2$ enables the resulting physics to be observed without the need for unity filling of sites. The reflectance spectrum is shown in Fig.\,\ref{b_setup}\,(d) for $N/2$ atoms with spacing $k\,a=\pi$, and differs significantly from the case of $N$ atoms spaced at $k\,a=\pi/2$.  Later, we will show that for the two-photon transition involving $\ket{s}$, Eq.\,(\ref{refl1}) holds but with re-scaled decay rates $\Gamma_{\rm 1D}$, $\Gamma'$.

\section{Energy shifts and effective decay rates}\label{sec_lshift}
As seen in the preceding section, the combination of band gap interactions ($\mathcal{H}_{\rm bg}$) and off-resonant driving ($\mathcal{H}_c$) leads to an effective energy shift of two-excitation states $\ket{e_m e_n}$ and $\ket{s_m s_n}$. The transition energies to these states determine the nonlinear response as seen by a probe field $\mathcal{E}_{\rm in}(z,t)$. We derive and explain these results in greater detail here, by deriving an effective Hamiltonian for the contribution $\mathcal{H}_{\rm es}=\mathcal{H}_{\rm bg}+\mathcal{H}_c$ in the framework of quasi-degenerate perturbation theory\,\cite{tannoudji04}.  
Such a description is valid in the dispersive limit of large control field detunings, here restricted to second order in the ratio of control field coupling strength and detuning $\sim |\Omega|/|\delta_c|$.

In the \textit{single excitation manifold}, the Hamiltonian $\mathcal{H}_{\rm es}$ is independent of the band gap coupling and takes on the simple form $ \mathcal{H}_{\rm es}^{(1)}=\delta_c\sum_m\,\ket{s_m}\bra{s_m} -\Omega\sum_m \bigl(\ket{e_m} \bra{s_m}+\mathrm{h.c.}\bigr)$,
where $\ket{j_m}=\sigma_{jg}^m\ket{g}^{\otimes N}$ denotes the state with atom $m$ excited to state $j$ and all other atoms in the ground state $\ket{g}$. In the limit $|\Omega |\ll|\delta_c|$ and to second order in $|\Omega|/|\delta_c|$ one obtains the effective form
\begin{equation}\label{heffst} \mathcal{H}_{\rm es}^{(1)}\simeq \left(-\frac{\Omega^2}{\delta_c} \right) \sum_m\ket{\tilde{e}_m} \bra{\tilde{e}_m}+\left(\delta_c+\frac{\Omega^2}{\delta_c}  \right)\,\sum_{m}\ket{\tilde{s}_m}\bra{\tilde{s}_m}  \end{equation}
with the dressed states defined as $\ket{\tilde{e}}=\ket{e}-\epsilon\,\ket{s}$ and $\ket{\tilde{s}}=\ket{s}+\epsilon\,\ket{e}$, where $\epsilon=\Omega/\delta_c$. 
Eq. (\ref{heffst}) thus describes the conventional (single-atom) ac-Stark shift, where a far detuned laser shifts the energies of the states involved in the transition by an amount $\pm \Omega^2/\delta_c$.

\begin{figure}[tbb]
\begin{centering}
\includegraphics[scale=0.52]{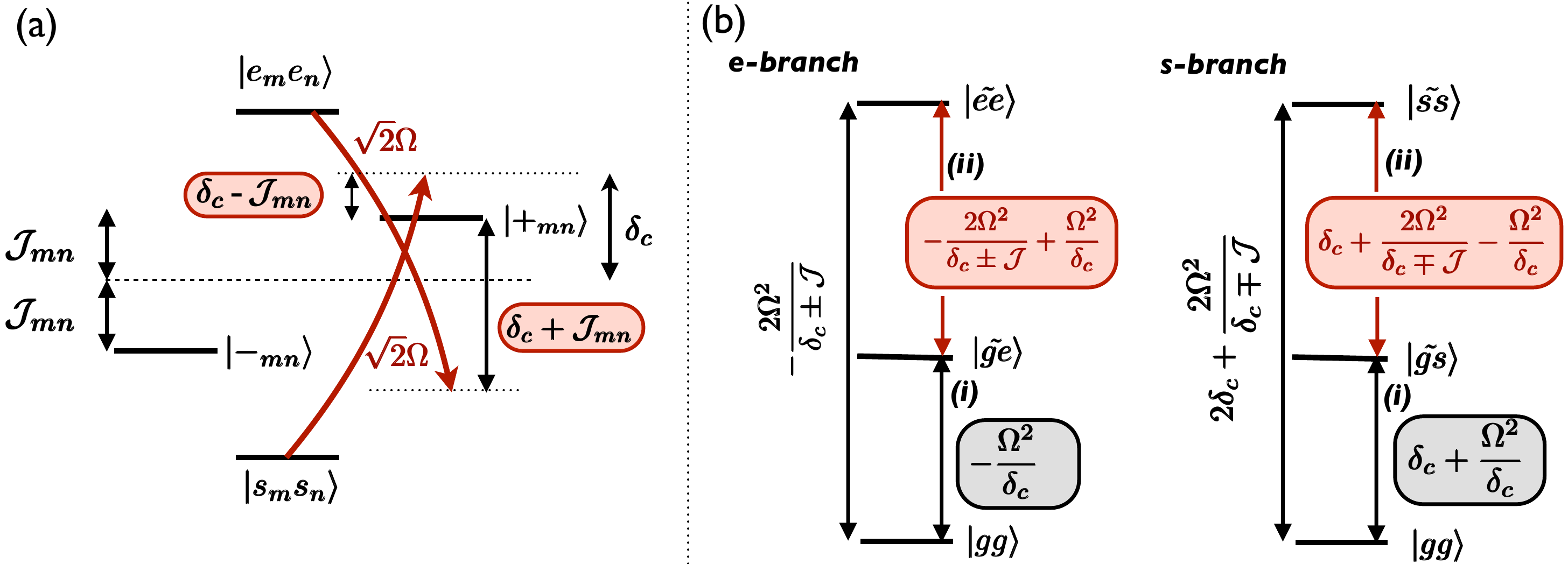}
\caption{\label{b_hameff} (a)  Energy levels and control field couplings and detunings in the two-excitation manifold. The band gap coupling $\mathcal{J}_{mn}=\mathcal{J}\,(-1)^{m+n}$ leads to a splitting of its eigenstates  $\ket{\pm_{mn}}=1/\sqrt{2}\,(\ket{e_ms_n}\pm\ket{s_me_n})$, resulting in an effective control field detuning $\delta_{\rm eff}^\pm=\delta_c\pm \mathcal{J}_{\rm mn}$ between the transitions from states $\ket{e_m e_n}$ and $\ket{s_m s_n}$. (b) Frequency separation for the $e$- (left) and $s$-branch (right) in the rotating frame and for $\mathcal{J}_{mn}=\pm \mathcal{J}$, indicating the frequency with respect to the probe field detuning $\delta$.  
}
\end{centering}
\end{figure}

In the \textit{two-excitation manifold}, the Hamiltonian $\mathcal{H}_{\rm es}$ can be written as
\begin{eqnarray}\label{h2c}\fl\quad\eqalign{\mathcal{H}_{\rm es}^{(2)}&=2\delta_c\sum_{(mn)} \ket{s_ms_n}\bra{s_ms_n} +\sum_{(mn)} \left(\delta_c+\mathcal{J}_{mn}\right)\,\ket{+_{mn}}\bra{+_{mn}} +\left( \delta_c-\mathcal{J}_{mn}\right)\,\ket{-_{mn}}\bra{-_{mn}}\\
  &-\sqrt{2}\Omega\,\sum_{(mn)}\left(\ket{e_m e_n} \bra{+_{mn}}+\ket{s_ms_n} \bra{+_{mn}}+\mathrm{h.c.}\right)}
\end{eqnarray}
where  $\ket{\pm_{mn}} =( \ket{e_ms_n}\pm\ket{s_me_n})/\sqrt{2}$ and the sum accounts for all possible two-excitation pair combinations.
Thus,  the control field effectively couples the states $\ket{s_ms_n}$ and $\ket{e_me_n}$ to $\ket{+_{mn}}$, with detunings $\delta_{\rm eff}^\pm=\delta_c\pm \mathcal{J}_{mn}$ (see Fig.\,\ref{b_hameff}\,(a)) that depend on the alternating sign of $\mathcal{J}_{\rm mn}$.

In the limit of $|\delta_c\pm\mathcal{J}|\gg |\Omega|$, the Hamiltonian to second order in $|\Omega|/|\delta_c\pm\mathcal{J}_{mn}|$ can be written as $\mathcal{H}_{\rm es}^{(2)}=\sum_{(mn)}\mathcal{H}_{mn}$ with
\begin{eqnarray}\label{eff_ham2}\fl \eqalign{\mathcal{H}_{ mn}\simeq & \frac{-2\,\Omega^2}{\delta_c+\mathcal{J}_{mn}}\ket{\widetilde{e_me_n}} \bra{\widetilde{e_me_n}}+\left(2\delta_c+\frac{2\,\Omega^2}{\delta_c-\mathcal{J}_{mn}}\right)\ket{\widetilde{s_ms_n}}\bra{\widetilde{s_ms_n}}+\left( \delta_c-\mathcal{J}_{mn}\right)\ket{-_{mn}}\bra{-_{mn}}\\ 
	&+\left(\delta_c+\mathcal{J}_{mn}+\frac{4\,\mathcal{J}_{mn}\,\Omega^2}{\mathcal{J}_{mn}^2-\delta_c^2}\right)\,\ket{\widetilde{+}_{mn}} \bra{\widetilde{+}_{mn}}
	-\left[\frac{2\,\Omega^2 \mathcal{J}_{mn}}{\mathcal{J}_{mn}^2-\delta_c^2}\,\ket{\widetilde{e_me_n}}\bra{\widetilde{s_ms_n}}+\mathrm{h.c.}\right]\,,}
	\end{eqnarray}
with the dressed states $\ket{\widetilde{ee}}=\ket{ee}+\sqrt{2}\,\epsilon_+\,\ket{+}$, 
$\ket{\tilde{+}}  = \ket{+}-\sqrt{2}\epsilon_+\,\ket{ee}-\sqrt{2}\,\epsilon_-\,\ket{ss}$, 
$\ket{\widetilde{ss}}=\ket{ss}+\sqrt{2}\,\epsilon_-\,\ket{+} $
to linear order in $\epsilon_{\pm}$ where $\epsilon_{\pm}=\Omega/(\mathcal{J}_{mn}\pm\delta_c)$.
In the limit $|\delta_c|\gg|\Omega|$ and $|\mathcal{J}|\lesssim |\delta_c|$ under consideration, actual transitions between dressed states, represented by the $\ket{\widetilde{ee}}\bra{\widetilde{ss}}$ term in (\ref{eff_ham2}), are negligible, and therefore the control field coupling leads to pure energy shifts.

We now analyze the transition energies for the $e$- and $s$-branch. Thereby the $e$-branch is defined as the transitions involving an excitation to $\ket{e}$ via a probe beam tuned near the atomic $g$ to $e$ transition ($\delta\simeq 0$). Analogously, the $s$-branch transitions are defined as the two-photon excitations to $\ket{s}$ and are addressed for detunings near a zero two-photon detuning ($\Delta\equiv \delta-\delta_c\simeq 0$). For the two  branches,  the corresponding transition frequencies needed to create one or two excitations -- effective detunings in the rotating Hamiltonian frame -- are depicted in Fig.\,\ref{b_hameff}\,(b).   The resonances for the probe field in that picture follow by choosing $\delta$ equal to the indicated frequencies. For the $e$-branch, and an atom-atom coupling $\pm\mathcal{J}$, this results in $\delta=-\Omega^2/\delta_c$ for the absorption of a single photon from the ground state, which simply represents the level shift arising from the ac-Stark shift induced by the control field. To absorb a second photon in the presence of an atomic excitation in $\ket{e}$ a detuning $\delta=-2\Omega^2/(\delta_c\pm \mathcal{J})+\Omega^2/\delta_c=-\Omega^2/\delta_c\pm(\Omega/\delta_c)^2(2\mathcal{J})+\mathcal{O}([\mathcal{J}/\delta_c]^2)$ is required, with the sign dependent on the corresponding atomic sublattice of double periodicity. Similarly for the $s$-branch, the single photon resonance follows as $\delta=\delta_c+\Omega^2/\delta_c$, and the resonance for a second excitation as $\delta=\delta_c+2\Omega^2/(\delta_c\mp\mathcal{J})-\Omega^2/\delta_c=\delta_c+\Omega^2/\delta_c\pm(\Omega/\delta_c)^2\,(2\mathcal{J})+\mathcal{O}([\mathcal{J}/\delta_c]^2)$.

 The linewidths of the transitions are determined by the effective coupling rates into the waveguide $\Gamma_{\rm 1D}^{\rm eff}$ and to free space $\Gamma'_{\rm eff}$, which follow from multiplying the amount of e-population in the corresponding dressed states with the bare rates $\Gamma_{\rm 1D}$ and $\Gamma'$. For the $e$-branch, and in the dispersive limit considered here, they are in good approximation given by the original rates, i.e. $\Gamma_{\rm 1D}^{\rm eff}\simeq \Gamma_{\rm 1D}$ and $\Gamma_{\rm eff}'=\Gamma'$. For the $s$-branch they follow as $\Gamma_{\rm 1D}^{\rm eff}\simeq (\Omega/\delta_c)^2\,\Gamma_{\rm 1D}$, $\Gamma_{\rm eff}'\simeq (\Omega/\delta_c)^2\,\Gamma'$ for the single excitation transition, and $\Gamma_{\rm 1D}^{\rm eff}\simeq \epsilon_-^2\,\Gamma_{\rm 1D}$, $\Gamma_{\rm eff}'\simeq \epsilon_-^2\,\Gamma'$ with $\epsilon_-= \Omega/(\mathcal{J}_{mn}-\delta_c)$ for the two-excitation transition.

Despite providing an intuitive qualitative picture, the model presented above is not sufficient to fully describe the transition frequency between the states $\ket{gs}$ and $\ket{ss}$ (see Fig.\,\ref{b_hameff}\,(b)). For the s-branch the linewidth  $\Delta\omega_{\rm res}\simeq (N/2)\, \Gamma_{\rm 1D}^{\rm eff}$ decreases at a faster rate $\sim (\Omega/\delta_c)^2$ than the increase of precision for the energy shifts and thus the resonance positions in the perturbative treatment.  Thus we cannot expect  our estimated resonance frequency to overlap (within less than the linewidth $\Delta\omega_{\rm res}$) with the full numerical solution, even though the essential properties are still well-approximated. 
A more accurate prediction of the resonance frequencies, which also provides excellent quantitative agreement for the $s$-branch, is given by a transfer-matrix model introduced in \ref{sec_tramodel}.

\section{Photon and excitation correlations}\label{sec_correl}
\begin{figure}[!t]
\begin{centering}
\includegraphics[scale=0.28]{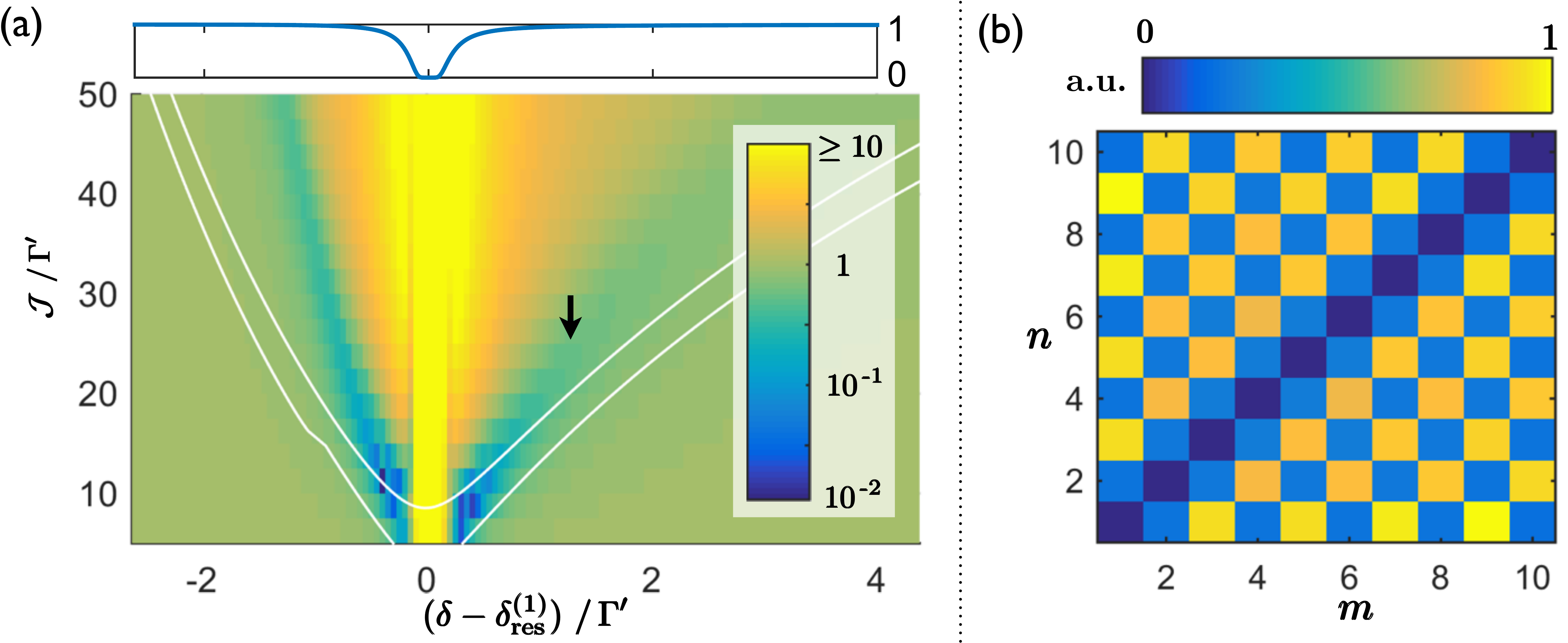}
\caption{\label{b_correl}  (a) Two-photon correlation $g^{(2)}(0)$ for the transmitted field vs relative field detuning $\delta-\delta_{\rm res}^{(1)}$ ($\delta_{\rm res}^{(1)}=97.62\,\Gamma'$) and atomic coupling rate $\mathcal{J}$. The overset illustrates the linear transmittance spectrum $\mathcal{T}^{(1)}=\ex{\mathcal{E}_+^\dagger \mathcal{E}_+}/\mathcal{E}_0^2$. White lines border the region of high reflectance ($\mathcal{R} >0.8$) for the case where an $s$-excitation is prepared deterministically. The numerical parameters used are $N=100$, $\Gamma_{\rm 1D}=0.5\,\Gamma'$, $\delta_c=94\,\Gamma'$, $\Omega=18.8\,\Gamma'$, $\mathcal{E}_0=10^{-4}\,\Gamma'$. (b) Wave-function population (arbitrary units) of the doubly excited states $\ket{s_ms_n}$  at the position marked by the black arrow ($\delta-\delta_{\rm res}^{(1)}=1.38\,\Gamma'$, $\mathcal{J}=25\,\Gamma'$) in (a). }
\end{centering}
\end{figure}

We now consider the effects of the change in band structure on a weak coherent state input.
Here and in the following we resonantly excite atoms to the `s-branch' by choosing a two-photon detuning $\Delta\sim 0$, and define $\delta=\delta_{\rm res}^{(1)}$ as the probe detuning where linear transmission is maximally attenuated.
 Signatures of nonlinearities are observable in two-photon correlations of the cw-probe field  $\tilde{\mathcal{E}}_{\rm in}(t)=\mathcal{E}_0$; in particular we illustrate  in Fig.\,\ref{b_correl}\,(a) $g^{(2)}(0)=\ex{\mathcal{E}_+^\dagger\mathcal{E}_+^\dagger\mathcal{E}_+\mathcal{E}_+}/\ex{\mathcal{E}_+^\dagger\mathcal{E}_+}^2$ of the transmitted probe field (see\,\ref{sec_numimp} for details of the calculation). 
We find two branches of photon antibunching $g^{(2)}(0)<1$, which are increasingly shifted away from the point of maximum linear attenuation $\delta=\delta_{\rm res}^{(1)}$ with increasing atomic coupling $\mathcal{J}$. The positions of the branches, which indicate a suppression of two-photon propagation, qualitatively follow the conditions for high reflectance of a second photon when a single atomic excitation is first prepared deterministically (white lines, calculated in detail later), however with a reduced splitting.
In the present case of a weak coherent state input, the spatio-temporal dynamics of multi-photon propagation appear quite complicated, and we cannot find a simple predictive model for the anti-bunching dips. However, we believe that the positions of the anti-bunching dips represent  a balance between efficient absorption of the first photon (maximal near $\delta_{\rm res}^{(1)}$), and decreased reflectance away from the resonances of a single sublattice (white lines).
While it might be difficult to observe experimentally, numerically we can find additional confirmation of the physics by examining the atomic populations.
Specifically, near the anti-bunching dips for $\delta > \delta_{\rm res}^{(1)}$, a checkerboard pattern, as the one illustrated for a selected point (black arrow of Fig.\,\ref{b_correl}\,(a)) in Fig.\,\ref{b_correl}\,(b),  is observed in the population of states $\ket{s_m s_n}$. Such a pattern reveals that indeed only atoms of a single sublattice are excited. 
For the opposite anti-bunching  branch  ($\delta < \delta_{\rm res}^{(1)}$) a complementary (inverted) pattern holds.

\section{Conditional single photon reflection}\label{sec_condphoton} 
We now consider the case where a ``gate" photon is initially and deterministically mapped onto a spin excitation, whose presence or absence controls the propagation of a subsequent  ``signal'' photon. Such a case is interesting because it removes the spatio-temporal complexity associated with a coherent state input, and also constitutes a logical quantum gate. The signal photon will be centered at a frequency corresponding to one of the sublattice resonances, which leads to strong reflection in the presence of the gate. We will also be primarily interested in the regime where the sublattice resonances are well-separated $\Delta\nu>(N/2)\,\Gamma_{\rm 1D}$. This implies that the signal photon is far off resonance from the atomic medium absent the gate and is thus nearly perfectly transmitted in that case. We address the regime of smaller resonance separations $\Delta\nu$ in \ref{sec_Jdependence}, where it is shown that suitable reflectance 
resonances can also be obtained in regimes of overlapping resonances, which drastically reduces the atom-atom coupling $\mathcal{J}$ required by around an order of magnitude.

\begin{figure}[tbh]
\begin{centering}
\includegraphics[scale=0.32]{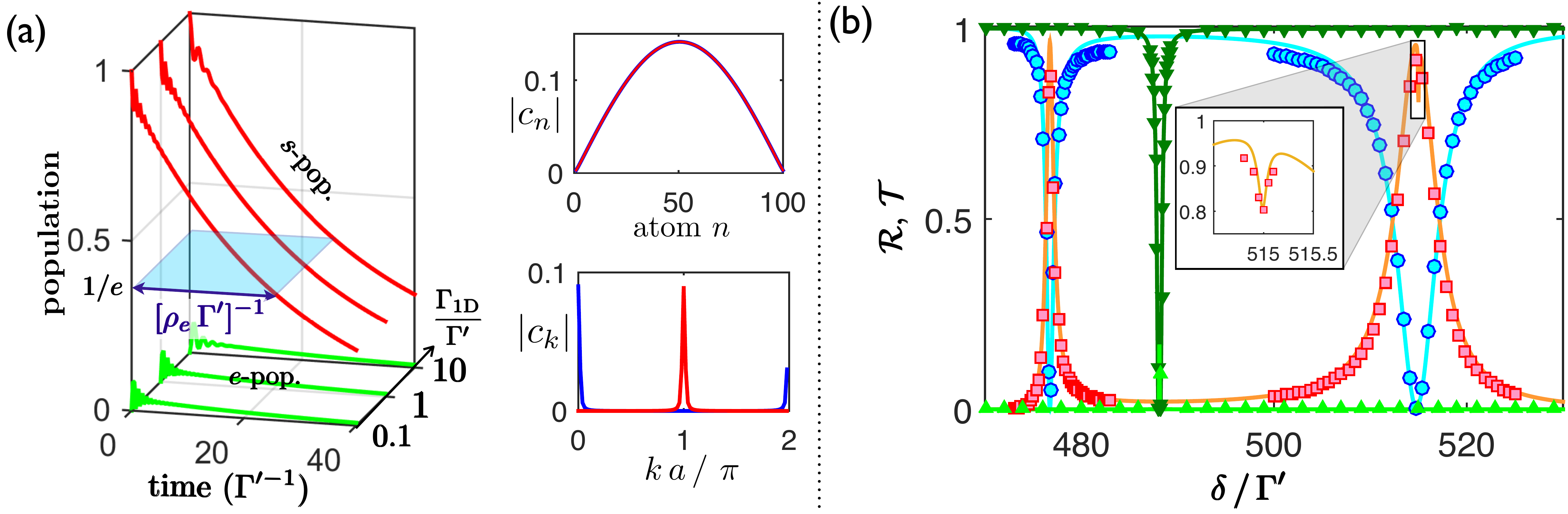}
\caption{\label{b_decspec} (a) Left: Atomic population decay in time for an initial subradiant excitation in $\ket{s}$, an atom number $N=100$, and for different ratios of waveguide $\Gamma_{\rm 1D}$ to free-space $\Gamma'$ decay. The $e$- and $s$-populations are defined as the total populations $p_e^{\rm tot}=\sum_{n=1}^{N} \ex{\sigma_{\rm ee}^{(n)}}$ and $p_s^{\rm tot}=\sum_{n=1}^{N} \ex{\sigma_{\rm ss}^{(n)}}$, respectively.  Right: Degenerate (blue and red) single excitation subradiant states  $\ket{\psi}=\sum_n c_n\,\exp(ik a\,n)\,\ket{s_n}$  for $N=100$. The upper plot illustrates the amplitude $|c_n|$ in the atom numbering $n$, the lower plot the amplitude in momentum space $|c_k|$ obtained by a discrete Fourier transform of $\ket{\psi}$. 
(b) Conditional reflectance $\mathcal{R}$ (red, squares) and transmittance $\mathcal{T}$ (blue, circles) spectra for the s-branch in the presence of a non-decaying gate excitation and $N\,\Gamma_{\rm 1D}/\Gamma'=100$. Green and light green lines (triangles) correspond to the transmittance and reflectance in the absence of an initial gate excitation, respectively. Data points represent numerical results, solid lines the transfer-matrix model predictions. \emph{Parameters: } $\delta_c=470\,\Gamma'$, $\Omega=94\,\Gamma'$, $\mathcal{J}=235\,\Gamma'$. }
\end{centering}
\end{figure}

The gate photon can be mapped onto a spin excitation $\ket{s}$ using, e.g., electromagnetically induced transparency (EIT). 
It results in a spin wave excitation $\ket{\psi}\sim \sum_n e^{i\,k_{\rm EIT}\,z_n} c_n\,\ket{s_n}$ with an error\,$\sim \Gamma'/(N\Gamma_{\rm 1D})$\,\cite{douglas15mol,fleischhauer05, gorshkov07}. 
 Here $k_{\rm EIT}=\pi/(2a)$ is the wavevector corresponding to the EIT transparency window, $c_n$ is a slowly-varying envelope that depends on the details of the initial pulse shape, and $\ket{s_n}\equiv \sigma_{\rm sg}^n\ket{g}^{\otimes N}$. Once the gate photon is mapped in, directly applying the control field $\Omega$ would cause the spin-wave to propagate and be mapped out into a photon. To ``trap'' this excitation inside the atomic gas, and allow an efficient interaction with the signal photon, one can first apply a spatially varying phase (e.g. by a magnetic field gradient) to shift the central wavevector, $k_{\rm EIT}\to 2\,k_{\rm EIT}$. With the appropriate amplitude shape, the resulting state can be made subradiant to the waveguide, such that it represents an eigenstate of $\mathcal{H}_{\rm wg}$ with minimal decay rate.

The form of such a subradiant state $\ket{\psi_d}=\sum_n c_n\,e^{ika\,n}\ket{s_n}$ is illustrated in Fig.\,\ref{b_decspec}\,(a) (right plots), where both the spatial amplitude $|c_n|$ and  the amplitude in momentum space $|c_k|$ are plotted. To define such a state, one first identifies the eigenstate $\ket{j}$ of $\mathcal{H}_{\rm wg}$ in the single excitation manifold $\{\ket{e_n} \}$, with minimal collective waveguide decay rate $\Gamma_{\rm wg}^{\rm (1D)}=-2\,{\rm Im}(\ex{j|\mathcal{H}_{\rm wg}|j})$. Out of that subradiant state the analogue one in states $\ket{s}$ is obtained by substituting $\ket{e_n}\to \ket{s_n}$. Two degenerate such states exist (blue and red line in Fig.\,\ref{b_decspec}\,(a), respectively), which are centered in momenta around $k=0$ and $k=2k_{\rm EIT}=\pi/a$. For these subradiant states the waveguide decay scales as $\Gamma_{\rm wg}^{\rm (1D)}\propto \rho_e \Gamma_{\rm 1D} N^{-3}$ with the atom number $N$. Here, $\rho_e\,\simeq (\Omega/\delta_c)^2$ is the amount of population in $\ket{e}$ that mixes with $\ket{s}$ due to the control field. 

 The decay of an initial subradiant excitation is shown in Fig.\,\ref{b_decspec}\,(a) (left plot) for different ratios of $\Gamma_{\rm 1D}/\Gamma'$. 
One generally observes that the decay of an excitation within the chain, apart from the initial time period, is essentially independent of the waveguide decay rate and given by $\Gamma_{\rm dec}\simeq \rho_e\,\Gamma'$.  More generally, the collective nature of $\mathcal{H}_{\rm wg}$ splits the single excitation manifold into a few states that radiate efficiently into the waveguide and a vast number of states with subradiant character. Thus, even though the subradiant state defined above represents the optimal choice, any state with $\Gamma_{\rm wg}^{\rm (1D)}<\Gamma'$, e.g. a product state with only a single atom $m$ excited to $\ket{s_m}$, reveals a comparable decay behavior dominated by the decay to free space $\Gamma'$.

Having established the fidelity for the gate photon to be mapped into a spin excitation and its lifetime, we now consider its effect on a signal photon. In order to separate the finite decay rate of the spin excitation arising from the gate photon from the propagation dynamics of the signal, we first consider a system where one atom is initialized in $\ket{s}$, and whose decay rates $\Gamma_{\rm 1D}=\Gamma'=0$ are set to zero. Thus it only sees the rest of the atomic chain through the band gap coupling.  The response to the signal field is then calculated by considering the system response to a weak cw-input field $\tilde{\mathcal{E}}_{\rm in}(t)=\mathcal{E}_0$. This results in a reflectance spectrum as shown by the data points in Fig.\,\ref{b_decspec}\,(b). The results also show good agreement with a simpler effectively linear optical model (solid lines). Here the reflection and transmission coefficients of individual atoms are alternated by hand, and the reflectance and transmittance of the entire array is obtained by a transfer matrix calculation, whose details are given in  \ref{sec_tramodel}.\\
In the limit $\Delta\nu>(N/2)\,\Gamma_{\rm 1D}$, the s-branch spectrum consists of two resonances ($\pm$) of Lorentzian shape. The response in reflection around these resonances can be approximated by $\mathcal{R}(\Gamma_{\rm 1D}^{\rm eff}, \Gamma'_{\rm eff}, \delta_{\pm})$ \,(\ref{refl1}). Here $\Gamma_{\rm 1D}^{\rm eff}=\epsilon_{-}^2\, \Gamma_{\rm 1D}$ and $\Gamma'_{\rm eff}=\epsilon_{-}^2\,\Gamma'$ are the effective waveguide and free space decay rates of the transition, respectively, and $\delta_{\pm}$ the resonance detuning. The parameter $\epsilon_{-}$ characterizes the amount of $e$\,-\,amplitude  mixed into $s$ in the two-excitation manifold, and  has been derived in section\,\ref{sec_lshift} to $\epsilon_{-}\simeq \Omega/(\pm\mathcal{J}-\delta_c)$.
A detrimental effect, associated with the off-resonant interactions between the optical field and the far-detuned atomic sublattice, and discussed in more detail in \ref{sec_crossinterf}, appears in the peak reflectance as interference dips (inset of Fig.\,\ref{b_decspec}\,(b)). These dips become more prominent for larger optical depth $\mathcal{D}=2N\Gamma_{\rm 1D}/\Gamma'$ and decreasing frequency difference between the sublattice resonances.
Aside from increasing the latter, they can also be suppressed by modifying the propagation phase of the bare waveguide $k\,a=\pi\,(1+\kappa)/2$, here characterized by a small adjustment parameter $\kappa$ which is specified in detail in \ref{sec_crossinterf}.

\begin{figure}[thb]
\begin{centering}
\includegraphics[scale=0.32]{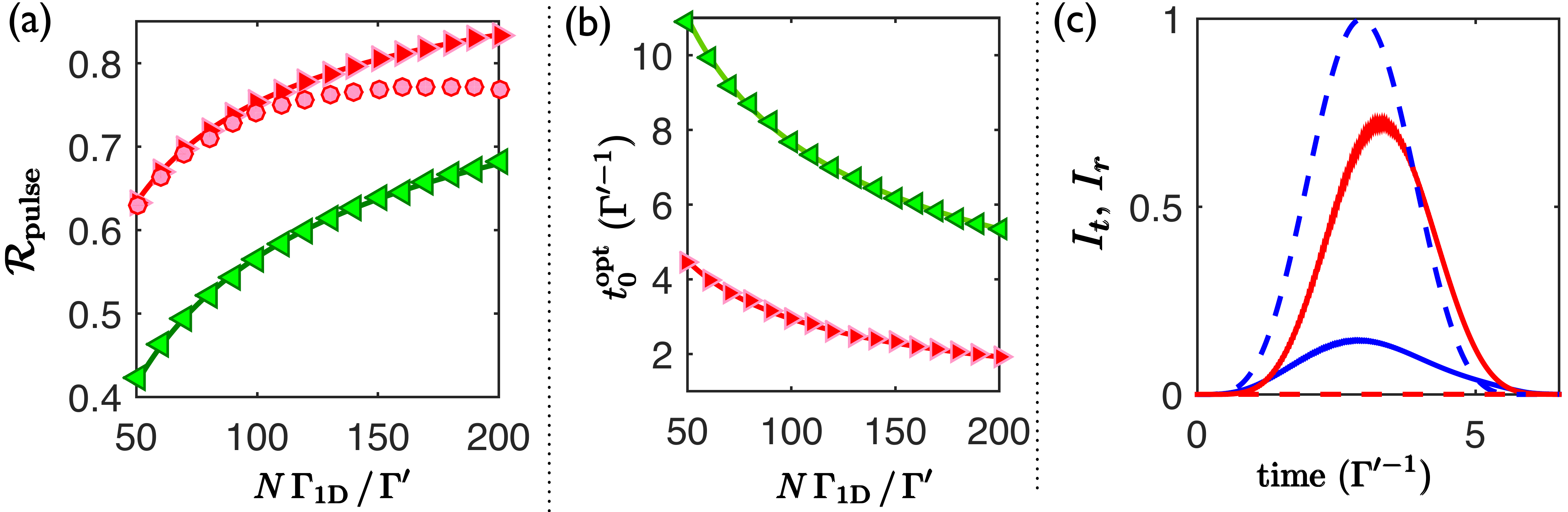}
\caption{\label{b_scaling}  (a) Signal pulse reflectance $\mathcal{R}_{\rm pulse}$ vs $N\Gamma_{\rm 1D}/\Gamma'$ for an initial gate excitation prepared at $t$=0 and a signal pulse $\tilde{\mathcal{E}}_{\rm in}(t)=\mathcal{E}_0\,\sin^2(\pi\,t/(2\,t_0))$ for $0\leq t\leq 2t_0$.  The temporal pulse width $t_0$ has been optimized, with the optimized values shown in (b). The red triangles and red circles correspond to the signal pulse resonant with the right (broader) s-branch resonance of Fig.\,\ref{b_decspec}\,(b) with phase adjustment parameter $\kappa=9\mbox{e-4}$ and $\kappa=0$, respectively. Green triangles account for the signal pulse resonant with the left (narrower) s-branch resonance ($\kappa=0$).  (c)
Reflected $I_r$ (red) and transmitted $I_t$ (blue) signal pulse intensity (normalized by $|\mathcal{E}_0|^2$) for $N\Gamma_{\rm 1D}/\Gamma'=100$ ($N$=200, $\Gamma_{\rm 1D}$=$\Gamma'/2$, $\mathcal{E}_0=\mbox{2e-4}\,\Gamma'$) on the right (broader) s-branch resonance with optimal pulse width $t_0=2.95\,\Gamma'^{-1}$. Solid and dashed lines correspond to the presence and absence of a gate excitation, respectively.   }
\end{centering}
\end{figure}

Having analyzed the gate excitation lifetime and the excitation-dependent reflectance properties independently, we now consider the full dynamics of a signal pulse interacting with a (decaying) initial gate excitation. In order to exploit the reflectance properties, the pulse time $t_0$ must be small compared to the gate excitation lifetime, and its bandwidth $\propto 1/t_0$ small compared to the reflectance resonance linewidth: $\Gamma_{\rm dec}<t_0^{-1}<\Delta\omega_{\rm res}$. 
We illustrate the optimal signal pulse reflectance $\mathcal{R}_{\rm pulse}$ and its scaling with $N\Gamma_{\rm 1D}/\Gamma'$ in Fig.\,\ref{b_scaling}\,(a), obtained for the numerically optimized pulse time $t_0^{\rm opt}$ depicted in Fig.\,\ref{b_scaling}\,(b). Here we assume the $s$-branch configuration as in Fig.\,\ref{b_decspec}\,(b) with an initial (and decaying) gate excitation in $s$ prepared at $t=0$, followed by the launch of a signal pulse of shape $\tilde{\mathcal{E}}_{\rm in}(t)=\mathcal{E}_0\,\sin^2(\pi\,t/(2\,t_0))$ ($0\leq t\leq 2\,t_0$, $\tilde{\mathcal{E}}_{\rm in}(t)=0$ otherwise). The reflectance $\mathcal{R}_{\rm pulse }$ is defined as the ratio of time-integrated reflected output to input intensity, where the background field leaked by the decaying excitation has been subtracted.

    As expected a larger ratio $\Delta\omega_{\rm res}/\Gamma_{\rm dec}$ improves the reflectance. Thus, centering the gate pulse around the right (broader) resonance of Fig.\,\ref{b_decspec}\,(b), yields a 
higher reflectance as compared to the left (narrower) resonance.
In Fig.\,\ref{b_scaling}\,(a), one also sees the effect of the interference dips. In particular, without compensating for them, the reflectance (red circles) saturates around $77\%$. In contrast, no saturation emerges in the compensated (propagation phase adjusted) version, reaching  $83\%$ reflection for $N\,\Gamma_{\rm 1D}/\Gamma'=200$ ($\mathcal{D}=400$). An example for the probe pulse propagation is illustrated in Fig.\,\ref{b_scaling}\,(c), comparing the propagation in the presence and absence of an initial excitation. A clear inversion from dominant reflection in the former to transmission in the latter case is observed.

\section{Influence of dephasing}\label{sec_dephasing}

Here we address the impact of dephasing on the conditional reflectance of a photon. The dephasing rates of atoms near photonic crystal waveguide structures are not known yet, but have been measured for a similar system consisting of atoms trapped in the close vicinity ($\sim$200 nm) of nanofibers. For these setups decoherence (dephasing) rates in the range of $\gamma\simeq 2\pi\,\left[200\,{\rm Hz} - 50\,{\rm kHz}  \right]$ have been reported\,\cite{reitz13, sayrin15} for cesium atoms, where the lowest rates correspond to the magnetically insensitive microwave clock transition\,\cite{reitz13}. Their origin has been attributed to temperature dependent ac-Stark shifts in the optical trap geometry.  In order to relate these quantities to the  free space $\Gamma'$  and waveguide emission rate $\Gamma_{\rm 1D}$, we assume that $\Gamma'=2\pi\cdot 4.56\,{\rm MHz}$ as reported for Cs atoms trapped along photonic crystal waveguides in\,\cite{hood16} and $\Gamma_{\rm 1D}\simeq \Gamma'$ as reported in\,\cite{goban15}. With the dephasing parameter range above, this results in $\gamma/\Gamma'\simeq\gamma/\Gamma_{\rm 1D} \simeq [4\cdot 10^{-5} - 0.01]$.

\begin{figure}[tbh]
\begin{centering}
\includegraphics[scale=0.5]{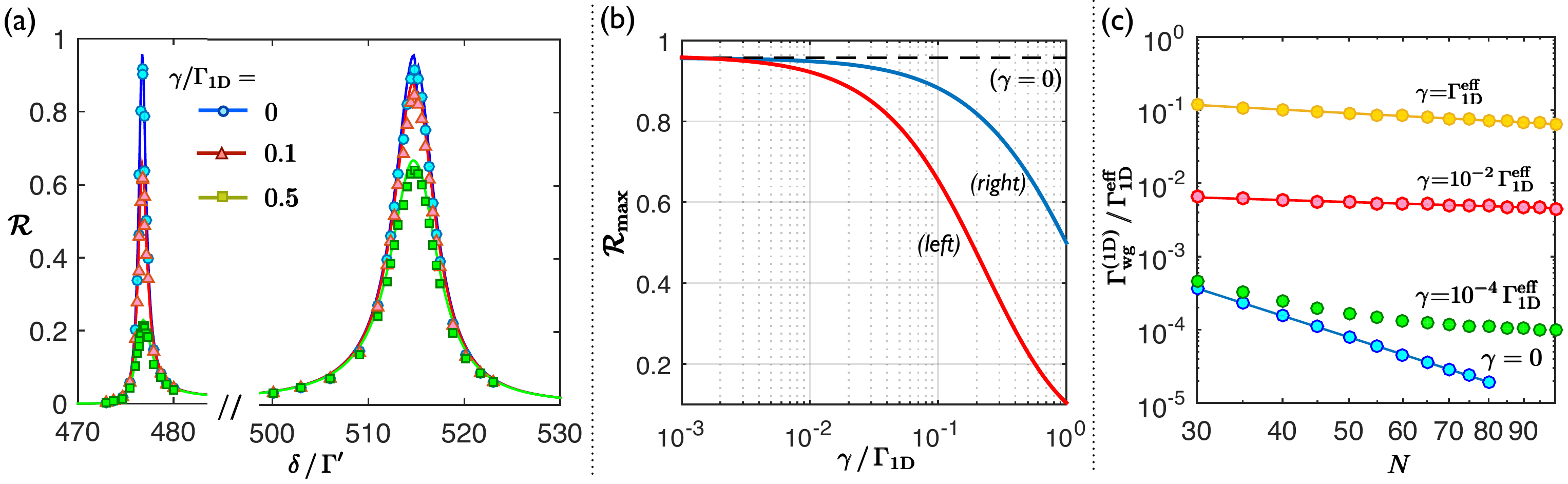}
\caption{\label{b_dephasing} 
(a) $s$-branch reflectance spectrum for different dephasing rates $\gamma$, with the parameters chosen as in Fig.\,\ref{b_decspec} (and $\Gamma_{\rm 1D}=\Gamma'$). Solid lines correspond to the transfer-matrix prediction, data points to a numerical simulation.  (b) Maximal reflectance amplitude $\mathcal{R}_{\rm max}$ for the setup as in (a) vs dephasing rate $\gamma$ for the left (red) and right (blue) reflectance resonance. The dashed line indicates the reflectance amplitude in the absence of dephasing.  (c) Waveguide decay rate $\Gamma_{\rm wg}^{\rm(1D)}$ for an initial subradiant state superposition vs atom number $N$ and for different dephasing rates $\gamma$ (double logarithmic scale). Solid lines are fits to a scaling $\Gamma_{\rm wg}^{\rm(1D)}\sim N^{-\alpha}$, except for the case of $\gamma=10^{-4}\Gamma_{\rm 1D}^{\rm eff}$, which clearly does not scale polynomially with $N$. } 
\end{centering}
\end{figure}

In the following we model dephasing as uncorrelated noise, which in the framework of a Markovian (Lindblad form) master equation can be described as ($\sigma_{mm}=\ket{m}\bra{m}$)
\begin{equation}\label{mdeph1} \mathcal{L}^{\rm deph}[\rho]=\sum_m\frac{\gamma_m}{2}\,\left(2\sigma_{\rm mm}\rho\,\sigma_{\rm mm}-\rho\,\sigma_{\rm mm}-\sigma_{\rm mm}\,\rho \right)  \end{equation}
resulting in a decoherence rate $\gamma_{mn}=(\gamma_m+\gamma_n)/2$ on the $m$ to $n$ transition. As $\gamma\ll\Gamma'$ it is sufficient to only explicitly consider dephasing on the $g$ to $s$ transition whereas additional dephasing affecting the excited state can be ignored, and thus the rates in (\ref{mdeph1}) are chosen such that only $\gamma_{gs}=\gamma\neq 0$.

Dephasing affects the conditional photon reflection at two stages: It modifies the reflectance spectrum as seen by a second `signal' photon (i.e. Eq.\,(\ref{refl1}) and Fig.\,\ref{b_decspec}\,(b)) and it reduces the lifetime of a first `gate' photon (Fig.\,\ref{b_decspec}\,(a)). 

The impact on the reflectance spectrum has been analyzed by re-deriving the transfer-matrix model as introduced in\,\ref{sec_tramodel} including dephasing of the form\,(\ref{mdeph1}). The validity of that approach is verified in Fig.\,\ref{b_dephasing}\,(a) by comparing the predicted reflectance to the one obtained by a fully numerical Hamiltonian evolution, in complete analogy to the procedure in the absence of dephasing (Fig.\,\ref{b_decspec}\,(b)). It illustrates that dephasing essentially reveals itself as a reduction of the reflectance amplitude. Based on the transfer-matrix model, the maximal reflectance amplitude $\mathcal{R}_{\rm max}$ dependence on the dephasing rate $\gamma$ is shown in Fig.\,\ref{b_dephasing}\,(b) for the two s-branch resonances of Fig.\,\ref{b_decspec}\,(b). This analysis suggests that the impact of dephasing becomes important for $\gamma/\Gamma_{\rm 1D}>10^{-2}$, whereas for the dephasing parameters stated above we do not expect significant amplitude reductions.

More generally, the influence of a dephasing $\gamma$ on the reflectance resonances can be incorporated into the reflectance formula\,(\ref{refl1}). Around a resonance $\delta=\lambda$, and assuming that $\gamma\ll|\lambda-\delta_c|$, the reflectance follows as ($\delta'=\delta-\lambda$)
\begin{equation}\label{rfdeph} \mathcal{R}(\delta')=\frac{p_\lambda^2\,[(N/2)\Gamma_{\rm 1D}]^2}{p_\lambda^2\,[(N/2)\Gamma_{\rm 1D}+\Gamma'+\gamma\,\kappa_\lambda]^2+(2\delta')^2}\,.  \end{equation}
Here $p_\lambda$ denotes the $e$-population of the underlying dressed state on resonance, which determines the effective decay rates $\Gamma_{\rm 1D}^{\rm eff}=p_\lambda\Gamma_{\rm 1D}$ and $\Gamma'_{\rm eff}=p_\lambda\,\Gamma'$, i.e. $p_\lambda=\epsilon_-^2$ for the s-branch and with the definitions of Section\,\ref{sec_lshift}. The coefficient $\kappa_\lambda$ increases with increasing $s$-type character of the resonance, and, in the limit when the atom-atom coupling is much smaller than the control field detuning $|\mathcal{J}|\ll|\delta_c|$, is given by
\begin{equation}\label{rfcoeff} \kappa_\lambda=1+\frac{2\Omega^2}{(\lambda-\delta_c)^2}\,.  \end{equation}
Formulas (\ref{rfdeph}) and (\ref{rfcoeff}) can be analytically derived out of the transfer matrix model for a chain of $N/2$ atoms with periodicity $k\,a=\pi$ and for $\mathcal{J}=0$. Moreover, (\ref{rfdeph}) has  been numerically verified beyond that limiting case; i.e. it also holds true for $|\mathcal{J}|\lesssim |\delta_c|$ but with a coefficient $\kappa_\lambda$ that must be numerically inferred. Thus, in regard to the reflectance spectrum, the effect of dephasing can be  folded into a modification of $\Gamma'$, i.e. substituting $\Gamma'\to\Gamma'+\gamma\,\kappa_\lambda$ with a correction $\gamma\,\kappa_\lambda$, the latter expected to be small for realistic dephasing parameters.

We now discuss the influence of dephasing on an initial `gate' excitation. Preparing such an excitation in a subradiant state configuration -- a coherent superposition -- serves to increase its lifetime, i.e. makes its collective decay into the waveguide $\Gamma_{\rm wg}^{\rm(1D)}\ll\Gamma_{\rm 1D}^{\rm eff}$ subradiant and the lifetime is then dictated by the free space decay rate $\Gamma'_{\rm eff}$ alone as seen in Section\,\ref{sec_condphoton}. In contrast, in the absence of coherence -- in the presence of strong dephasing -- $\Gamma_{\rm wg}^{\rm (1D)}=\Gamma_{\rm 1D}^{\rm eff}$ and the total decay rate is given by the combination of $\Gamma'_{\rm eff}+\Gamma_{\rm 1D}^{\rm eff}$. Importantly, conditional photon reflection in the proposed setup only depends on the presence, i.e. the  lifetime, of an excitation, but not on the specific state or the coherence of that excitation.
Note that the fact that the subradiant state decays into free space at a rate $\Gamma_{\rm eff}'$ even without dephasing is much different than the typical case of photon storage, where a pulse would be stored indefinitely. In particular, here, the stored excitation (in state $\ket{s}$) must be used to virtually populate state $\ket{e}$ via the control field $\Omega$, which enables an interaction with a second incoming photon. A weak dephasing rate, such as expected from the previously quoted experimental values, is thus expected to have negligible effect on top of the already important free-space emission $\Gamma'_{\rm eff}$. 
  
The waveguide decay rate $\Gamma_{\rm wg}^{\rm (1D)}$ and its scaling with the atom number $N$ for an initially prepared subradiant state is analyzed in Fig.\,\ref{b_dephasing}\,(c) for different magnitudes of the dephasing rate $\gamma$. It has been defined as the inverse 1/e-decay time. The latter is obtained out of a numerical simulation, performed based on a two-level system evolution under the full master equation with waveguide decay rate $\Gamma_{\rm 1D}^{\rm eff}$ and $\Gamma'=0$. Dephasing leads to two effects: It increases $\Gamma_{\rm wg}^{\rm (1D)}$ and it changes the particle number scaling, which in the absence of dephasing is given by $\Gamma_{\rm wg}^{\rm (1D)}\sim 1/N^3$. For an $s$-excitation as considered here, $\Gamma_{\rm 1D}^{\rm eff}\simeq (\Omega/\delta_c)^2\,\Gamma_{\rm 1D}\simeq 0.04\Gamma_{\rm 1D}$ and analogously $\Gamma'_{\rm eff}\simeq 0.04\,\Gamma'$. As seen from Fig.\,\ref{b_dephasing}\,(c), for realistic dephasing rates $\Gamma_{\rm wg}^{\rm (1D)}$ remains subradiant, thus $\Gamma_{\rm wg}^{\rm (1D)}<\Gamma'_{\rm eff}$ and the excitation lifetime essentially remains unaltered by dephasing.

 In conclusion, only minor modifications due to dephasing are expected for both the reflectance spectrum and the `gate' excitation lifetime and thus also for the conditional reflectance of photons. The main limiting mechanism is given by the emission into free space; for the reflectance dephasing can be ascribed to a (small) correction to that rate.

\section{Conclusions}\label{sec_conclusion}
In summary, we have proposed a system, in which a single excitation (`gate photon') controls the propagation of a subsequent photon by effectively modifying the topology of atoms along a PCW. In particular, its presence manifests itself as an alternation of atomic energy levels, such that subsequent photons can be made to `see'  only every other atom. This property has a tremendous impact on the propagation of photons, e.g.,  switches the system from being transmissive to highly reflective in the absence and presence of a gate excitation, respectively.
We have shown how this can be observed in two-photon correlations, where conditional reflectance reveals itself as anti-bunching. 

Our proposed mechanism involves a straightforward way to both create and probe a photon nonlinearity of alternating type within a single platform, a combination that makes the PCW setup unique and enables the access to novel types of nonlinearities. 
The consequences of such an alternating nonlinearity for many photons are yet unknown, and might inspire the active search for the creation and understanding of many-body states of light.
 Moreover, while we have only considered the limit of infinite-range interactions ($L\rightarrow\infty$) here, it would be interesting in the future to explore finite-range interactions, where a photon only changes the topology of its immediate surroundings.  It would also be interesting to consider analogous effects in higher dimension\,\cite{tudela15}.

\ack
 The authors thank H.J. Kimble and J.S. Douglas for useful discussions. The work was supported by the ERC Starting Grant FOQAL, the MINECO Plan Nacional Grant CANS, the Spanish Ministry of Economy and Competitiveness through the ``Severo Ochoa'' Programme for Centres of Excellence in R\&D (SEV-2015-0522), the Fundacio Cellex, and the CERCA Programme\,/\,Generalitat de Catalunya.

\appendix

\section{Transfer matrix model for the reflectance and transmittance spectra}\label{sec_tramodel}
In  Fig.\,\ref{b_decspec}\,(b) we have illustrated the reflectance and transmittance spectra, both in the presence and absence of a `gate' excitation in the system. It has been shown that the spectra obtained by a fully numerical simulation can be well-approximated by a linear optical transfer-matrix model. Here we outline in detail how that transfer-matrix model calculation is performed. 
\begin{figure}[htb]
\begin{centering}
\includegraphics[scale=0.55]{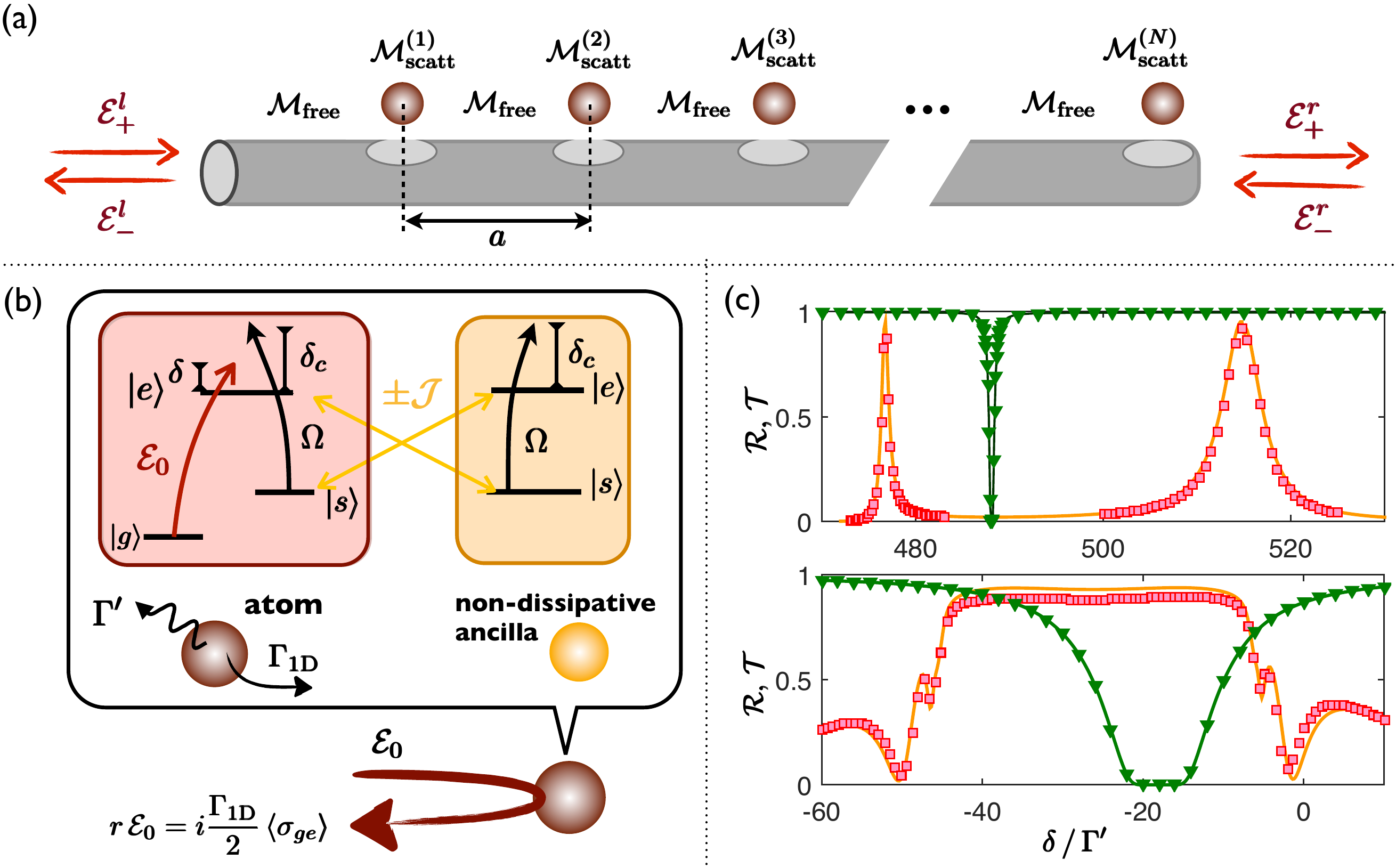}
\caption{\label{b_tmodel1} (a) Transfer matrix description for the field calculation. The input and output fields $\mathcal{E}$ are connected by a series of matrices $\mathcal{M}$ representing the scattering at an atomic site $\mathcal{M}_{\rm scatt}$ and free propagation sections $\mathcal{M}_{\rm free}$ in-between. (b) In the presence of a `gate' excitation, the scattering at a single atomic site is modeled by an atom coupled to a non-dissipative ancilla via the atom-atom interaction $\pm\mathcal{J}$. The atomic steady state coherence $\ex{\sigma_{\rm ge}}=\ex{\sigma_{\rm ge}^{\rm atom} \otimes \mathbbm{1}_{\rm ancilla}}$ determines the reflection coefficient $r$ at the atomic site. (c)
 Reflectance (red squares) in the presence of a gate excitation, and transmittance (green triangles) in its absence for the field tuned close to two-photon ($\delta-\delta_c\sim 0$) resonance ($s$-branch, upper panel) and single-photon ($\delta\sim 0$) resonance ($e$-branch, lower panel). In the former case, conditional reflectance arises from transitions to states $\ket{ss}$, in the latter case from transitions to states $\ket{ee}$.
The transfer matrix model predictions are given by the solid orange and green line, respectively.   \textit{Parameters: } $N=100$, $\delta_c=470\,\Gamma'$, $\Omega=94\,\Gamma'$, $\mathcal{J}=235\,\Gamma'$. }
\end{centering}
\end{figure}

For the propagation of photons in one dimension, a very elegant way to obtain linear reflectance and transmittance spectra is given by the transfer matrix approach\,\cite{chang12, deutsch95}. In that model, the propagation of photons is modeled by free propagation sections intersected by a series of scattering events at the (assumed point-like) atomic positions (see Fig.\,\ref{b_tmodel1}\,(a)). The input and output fields are then related by two-dimensional matrices $\mathcal{M}$
\begin{equation}\fl\left(\begin{array}{*{20}{c}}\mathcal{E}_+^{r} \\ \mathcal{E}_-^{r} \end{array}\right) = \mathcal{M}_{\rm tot}\left(\begin{array}{*{20}{c}} \mathcal{E}_+^l\\\mathcal{E}_-^l \end{array}\right)  \quad \mathrm{with}\quad \mathcal{M}_{\rm tot}=\mathcal{M}_{\rm scatt}^{(N)}\,\mathcal{M}_{\rm free}\,\mathcal{M}_{\rm scatt}^{(N-1)}\dots\mathcal{M}_{\rm scatt}^{(1)}\mathcal{M}_{\rm free} \end{equation}
where $+$ and $-$ denote the right (`transmitted') and left (`reflected') propagating field and $l$ and $r$ denote the left input (beginning of atom chain) and right output (chain end) position, respectively. The scattering matrices for atoms $1$ to $N$ are of the form
\begin{equation}\label{mscatt} \mathcal{M}_{\rm scatt}=\frac{1}{t}\left(\begin{array}{*{20}{c}} t^2-r^2 & r \\ -r & 1  \end{array}\right)   \end{equation}
with $t$ and $r$ being the frequency-dependent reflection and transmission coefficients associated with a single atom. The free propagation of distance $a$  imprints a pure phase and is described by
\begin{equation}\label{mfree} \mathcal{M}_{\rm free}=\left(\begin{array}{*{20}{c}}e^{ika} &  0\\ 0 & e^{-ika}  \end{array}\right)\,.
\end{equation}
In principle $M_{\rm free}$ is also frequency-dependent, through the frequency dependence of the wavevector $k$. Neglecting that dependence, i.e., evaluating $k$ at the atomic transition frequency, corresponds to a Markov approximation, which holds true for realistic atom numbers\,\cite{chang12, guimond16}.
For our system under consideration, we thus have $ka=\pi/2$ or $k\,a=(\pi/2)\,(1+\kappa)$ with a small adjustment $\kappa\ll 1$ (see \ref{sec_crossinterf}).
The total reflected and transmitted field coefficients then follow as $r_{\rm tot}=\mathcal{M}_{\rm tot}^{12}/\mathcal{M}_{\rm tot}^{22}$ and $t_{\rm tot}=1/\mathcal{M}_{\rm tot}^{22}$, respectively, with the numerical indices referring to the matrix elements of $\mathcal{M}_{\rm tot}$. The reflectance (transmittance), defined as the ratio of reflected (transmitted) to input field intensity is then given by $\mathcal{R}=|r_{\rm tot}|^2$ ($\mathcal{T}=|t_{\rm tot}|^2$).

Thus it remains to calculate the transmission and reflection coefficients $t$ and $r$. They can be obtained by solving for the steady state of the spin-model Hamiltonian $\mathcal{H}$ for a single atom. Based on the input-output relations\,\cite{chang12, caneva15} (see also Section\,\ref{sec_config}) $	\mathcal{E}_{+} =\mathcal{E}_0+i\,\frac{\Gamma_{\rm 1D}}{2}\,\sigma_{\rm ge}$ and $\mathcal{E}_-=i\,\frac{\Gamma_{\rm 1D}}{2}\,\sigma_{\rm ge}$ and noting that $\mathcal{E}_+=t\,\mathcal{E}_0$, $\mathcal{E}_-=r\,\mathcal{E}_0$ with the input field $\mathcal{E}_0$, the coefficients follow as
\begin{equation}\label{coeff1} r=i\,\frac{\Gamma_{\rm 1D}}{2}\frac{1}{\mathcal{E}_0}\,\ex{\sigma_{\rm ge}}^{[\rm 1st]}, \quad t=1+r \,.  \end{equation}
Here $\ex{\sigma_{\rm ge}}^{[\rm 1st]}$ denotes the steady state expectation value to first (linear) order in $\mathcal{E}_0$.

In the \emph{absence of an initial `gate' excitation} the atomic scattering matrices are equal, i.e. $\mathcal{M}_{\rm scatt}^{(k)}\equiv\mathcal{M}_{\rm scatt}$. The transmission and reflection coefficients follow out of
\begin{equation}\label{hamlinear}\fl  \mathcal{H}_{\rm atom}=-\left(\delta+i\frac{\Gamma'+\Gamma_{\rm 1D}}{2}\right)\,\sigma_{\rm ee}-(\delta-\delta_c)\,\sigma_{\rm ss}-\Omega\,(\sigma_{\rm es}+\mathrm{h.c.})-\mathcal{E}_0\,(\sigma_{\rm eg}+\mathrm{h.c.})\end{equation}
and (\ref{coeff1}), and for the three-level atom are given by the known coefficients\,\cite{chang11}
\begin{equation}\label{refl3} r=-\frac{\Gamma_{\rm 1D}\,(\delta-\delta_c)}{(\Gamma_{\rm 1D}+\Gamma'-2 i\,\delta)\,(\delta-\delta_c)+2i \,\Omega^2}\quad \mathrm{and }\,\, t=1+r\,. \end{equation}

A priori, the case where a \emph{`gate' excitation is present} seems beyond what can be captured by the transfer matrix approach for linear optics, since formally the response to a subsequent signal photon involves the atomic two-excitation manifold. However, we can construct a simpler artificial model (see Fig.\,\ref{b_tmodel1}\,(b)), in which the `gate' excitation is held in an ancilla atom that is decoupled from the waveguide -- thus it does not decay nor do photons in the waveguide directly excite this ancilla. In this model, the ancilla atom interacts via a photonic band gap, i.e. via the alternating atom-atom coupling $\pm\mathcal{J}$, to an `actual' atom that does couple to the waveguide. We can therefore calculate the linear optical properties of the actual atom, which is modified by the ancilla, and we find that our artificial model captures well the dynamics of the full system of interest.  
In that case, $\mathcal{M}_{\rm scatt}^{(2k)}=\mathcal{M}_{\rm scatt}^{[-\mathcal{J}]}$ and $\mathcal{M}_{\rm scatt}^{(2k+1)}=\mathcal{M}_{\rm scatt}^{[+\mathcal{J}]}$ for $k\in \mathds{N}_0$ with $\pm \mathcal{J}$ referring to the sign of the alternating atom-atom coupling constant. 
 The Hamiltonian is described by $\mathcal{H}=\mathcal{H}_{\rm atom}+\mathcal{H}_{\rm ancilla}+\mathcal{H}_{\rm int}$, denoting the system, two-level ancilla and interaction Hamiltonian, respectively. The atom (system) part $\mathcal{H}_{\rm atom}$ is given by\,(\ref{hamlinear}), the ancilla Hamiltonian takes the form
\begin{equation} \mathcal{H}_{\rm ancilla} =-\delta\,\sigma_{\rm ee}^{(2)}-(\delta-\delta_c)\,\sigma_{\rm ss}^{(2)}-\Omega\,(\sigma_{es}^{(2)}+\mathrm{h.c.})\end{equation}  
and the atom-atom interaction between both reads $\mathcal{H}_{\rm int}=\pm\mathcal{J}\,\sigma_{es}^{(1)}\otimes\sigma_{se}^{(2)}+\mathrm{h.c.}$. Here the indices $(1)$ and $(2)$ denote the atom and ancilla, respectively.
The reflection and transmission coefficients are obtained out of (\ref{coeff1}), by solving for the steady state expectation value $\ex{\sigma_{\rm ge}}^{[\rm 1st]}=\ex{\sigma_{\rm ge}^{(1)}\otimes \mathbbm{1}^{(2)}}^{[\rm1st]}$ to first (linear) order in $\mathcal{E}_0$. Thereby the initial excitation (in $e$ or $s$) is set by setting the populations in the coupled optical Bloch equations to the values
\begin{eqnarray}\eqalign{
 	\ex{\ket{g}_1\bra{g}\otimes \ket{e}_2\bra{e}}	&=1    	\quad \mbox{(e-branch)}\\
 	\ex{\ket{g}_1\bra{g}\otimes \ket{s}_2\bra{s}}	&=1    	\quad \mbox{(s-branch)}
}\end{eqnarray}
and all remaining populations to zero.   
Setting the populations to numerical values in the equations explicitly assumes them to be constant to a good approximation. This limits the validity of the formula to $(\Omega/\delta_c)^2\ll1$, i.e. to a regime in which the bare states $\ket{e}$ and $\ket{s}$ form a good basis of the coupled system. Moreover the population exchange between system and ancilla must be negligible, that is the approach is not valid for a signal field excitation to state $s$ with an initial (ancilla) `gate' $e$-excitation and vice versa. 

In Fig.\,\ref{b_tmodel1}\,(c) the transfer matrix model predictions for the reflectance and transmittance spectra (solid lines) are compared to the results of a full numerical simulation (data points) of the spin-model Hamiltonian.
The latter is obtained by numerically solving for the steady state reflectance and transmittance for a cw-input field (see also the upcoming\,\ref{sec_numimp}), where the initial state is set to all atoms being in the ground state (no initial `gate' excitation) or to an excitation of a single atom to $\ket{e}$ or $\ket{s}$ whose decay rates $\Gamma', \,\Gamma_{\rm 1D}$ are set to zero (initial `gate' excitation).  
Good agreement of the approaches is obtained, even when the parameters are not chosen to reside deep in the dispersive regime, e.g. $\Omega/\delta_c\simeq 0.2$, $\mathcal{J}/\delta_c\simeq 0.5$.

\section{Reflectance dependence on the band gap interaction parameter $\mathcal{J}$}\label{sec_Jdependence}
In Fig.\,\ref{b_decspec}\,(b) of Section\,\ref{sec_condphoton},  a large atom-atom coupling parameter $\mathcal{J}$ (as defined in (\ref{hambg})) has been chosen ($\mathcal{J}=235\,\Gamma'$) in order to achieve a large relative splitting between the two reflectance resonances.  Only in such a regime are the two resonances of the Lorentzian form\,(\ref{refl1}). Moreover, in that case the reflectance resonance frequencies are shifted away significantly from the resonances in the absence of a gate excitation, such that perfect transmission at these frequencies is obtained in the latter case. By partly relaxing at least one of these two requirements, we here show that it is possible to obtain reflectances of comparable linewidth even for signicantly lower coupling parameters $\mathcal{J}$.

\begin{figure}[htb]
\begin{centering}
\includegraphics[scale=0.55]{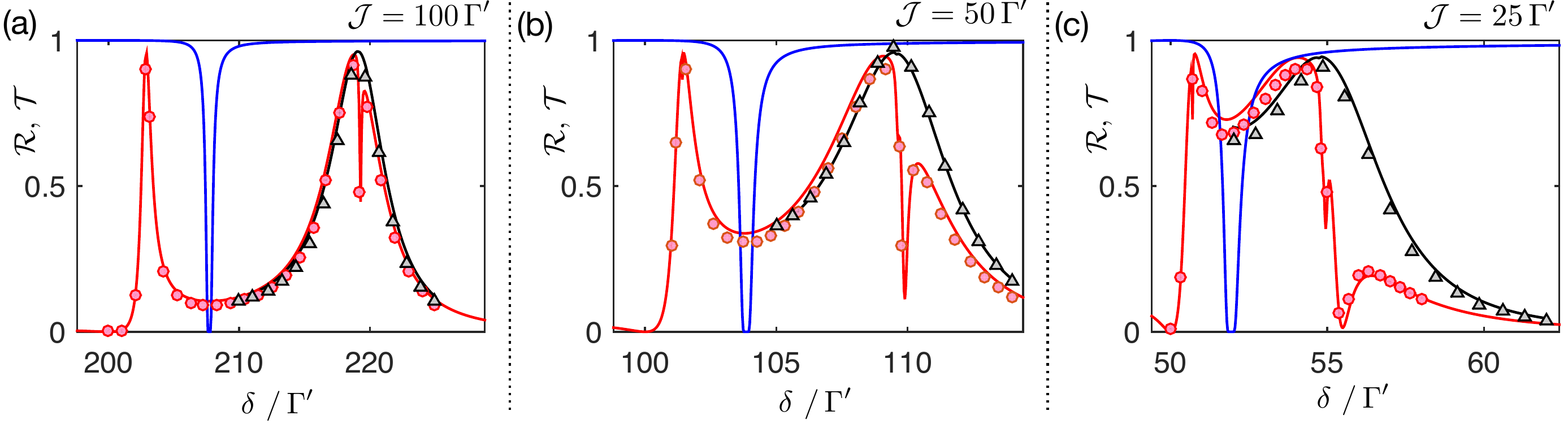}
\caption{\label{b_Jspect} 
Reflectance for different band gap coupling strengths $\mathcal{J}$. Here, the ratios $\Omega/\delta_c=0.2$ and $\mathcal{J}/\delta_c=0.5$ are kept constant and correspond to the ones in Fig.\,\ref{b_decspec}\,(b). Red lines and circles depict the reflectance $\mathcal{R}$ in the presence of a gate excitation, blue lines the transmittance $\mathcal{T}$ in the absence of the latter. Black lines and triangles correspond to an optimization of the right resonance by the phase adjustment parameters (a) $\kappa=2\cdot 10^{-3}$, (b) $\kappa = 4\cdot 10^{-3}$ and (d) $\kappa =8.5\cdot 10^{-3}$. Solid lines are obtained out of the transfer matrix model, data points (circles, triangles) correspond to  numerical data points obtained as described in Section\,\ref{sec_condphoton}.   } 
\end{centering}
\end{figure}

In Fig.\,\ref{b_Jspect}, reflectance spectra (red lines, in the presence of a gate excitation) are shown for different magnitudes of $\mathcal{J}$. The ratios of $\Omega/\delta_c=0.2$ and $\mathcal{J}/\delta_c=0.5$ are kept fixed and correspond to the ones in Fig.\,\ref{b_decspec}. This implies that both the excitation lifetime Fig.\,\ref{b_decspec}\,(a) and the reflectance linewidths -- up to disturbances and overlaps -- remain unaltered. For decreasing $\mathcal{J}$ the two resonances overlap, and finally form a common resonance around the resonance absent a gate excitation (dip position of the blue lines, which represent the transmission without any prior excitations). Moreover, disturbances and reflectance dips due to interference with the off-resonant sub-lattices become more prominent for smaller $\mathcal{J}$ and thus smaller resonance separations. However, this latter effect can be compensated for by adjusting the phase factor $k\,a =\pi/2\,(1+\kappa)$ with $\kappa$ a suitable correction factor, as further discussed in\,\ref{sec_crossinterf}. Such an adjustment works very well to reconstruct one of the resonances (black lines, here performed for the right resonance), even in regimes of large overlaps and disturbances (e.g., $\mathcal{J}=25\,\Gamma'$). Thus, suitable reflectance resonances can be constructed even for coupling parameters $\mathcal{J}$ that are an order of magnitude lower than in the well-separated Lorentzian configuration of Fig.\,\ref{b_decspec}\,(b). 

\section{Cross-interference reflectance dips and their compensation}\label{sec_crossinterf}
An atomic lattice with lattice constant $ka=\pi$ ideally produces a Lorentzian reflectance spectrum as described by\,(\ref{refl1}). Then in the case of band-gap interactions, we argued that an array of atoms with lattice constant $ka=\pi/2$ effectively appears as two sub-lattices each of lattice constant $ka=\pi$ for a signal field, when a `gate'  excitation is first stored. For low atom number, the reflectance near the resonance of one sub-lattice fits well with a Lorentzian. For larger atom number, however, the off-resonant response of the other sublattice cannot be ignored, and results in a significant interference dip in the reflectance (see inset of Fig.\,\ref{b_decspec}\,(b)). Here we analyze the origin of the reflectance dip in the transfer matrix formalism introduced in \ref{sec_tramodel}, and discuss how it can be eliminated by a slight change in the inter-atomic propagation phase, $ka=(\pi/2)(1+\kappa)$.
A repetitive unit in the transfer matrix description consists of a scattering event at the (quasi-) resonant atom $\mathcal{M}_{\rm scatt}$\,(\ref{mscatt}), followed by a free propagation $\mathcal{M}_{\rm free}$\,(\ref{mfree}) of distance $a$, an off-resonant scattering $\mathcal{M}_{\rm scatt}^{\rm off}$  (at the off-resonant sublattice), and another free propagation section $\mathcal{M}_{\rm free}$. 
For weak probe fields and large detunings, the individual off-resonant $e$- and $s$-transitions contribute independently and the scattering matrix can be written as $\mathcal{M}_{\rm scatt}^{\rm off}\simeq\mathcal{M}_{\rm scatt}^e\,\mathcal{M}_{\rm scatt}^{s}$ with $\mathcal{M}_{\rm scatt}^\mu$ the two-level scattering matrix of transition $\mu$. 
The two-level reflection and transmission coefficients are given by Eq.\,(\ref{refl3}) with $\Omega=\delta_c=0$, $\Gamma_{\rm 1D}\to\Gamma_{\rm 1D}^\mu$, $\Gamma'\to\Gamma'_\mu$ and $\delta\simeq\Delta_\mu$. Here $\Gamma_{\rm 1D}^\mu$ and $\Gamma'_\mu$ correspond to the effective decay rates of transition $\mu$.
The detuning $\Delta_\mu$ denotes the field detuning to the off-resonant sub-lattice resonance $\mu$. As we assume the probe field to be tuned to a particular reflectance resonance, whose reflectance dip behavior we seek to analyze, $\Delta_\mu$ simply corresponds to the reflectance peak separation.
Out of (\ref{mscatt}) and for large detunings $\Delta_\mu$, that is, to first order in $(\Gamma'+\Gamma_{\rm 1D})/\Delta_\mu$, this results in
\begin{equation}\label{mscattoff} \mathcal{M}_{\rm scatt}^{\mu}\simeq \left(\begin{array}{*{20}{c}} 1-i\,\kappa_\mu\,\frac{\pi}{2}  & -i\kappa_\mu\frac{\pi}{2} \\ i\kappa_\mu\,\frac{\pi}{2} & 1+i\kappa_\mu\,\frac{\pi}{2}  \end{array}\right)  \end{equation}
where we defined $\Gamma_{\rm 1D}^\mu/(2\,\Delta_\mu)=\kappa_\mu\,\pi/2$.

The idea is to compensate for the disturbance introduced by $\mathcal{M}_{\rm scatt}^{\mu}$ through adjusting the free space propagation phase as $k\,a=\pi\,(1+\kappa)/2$. Introducing that phase into the free propagation matrix $\mathcal{M}_{\rm free}$ (\ref{mfree}) results in
\begin{equation} \mathcal{M}_{\rm free}\simeq\left(\begin{array}{*{20}{c}}  e^{i\frac{\pi}{2}}\,(1+i\kappa\,\frac{\pi}{2}) & 0 \\ 0 &  e^{-i\frac{\pi}{2}}\,(1-i\,\kappa\,\frac{\pi}{2}) \end{array}\right) \end{equation}
where for consistency the adjustment parameter has been expanded to first order.  

A single repetitive unit thus follows as $\mathcal{M}_{\rm unit}=\mathcal{M}_{\rm free}\,\mathcal{M}_{\rm scatt}^{\rm off}\,\mathcal{M}_{\rm free}\,\mathcal{M}_{\rm scatt}$.
Note that if one removes the contribution of the off-resonant sub-lattice (which can be achieved by setting $\kappa_\mu=0$ in $M_{\rm scatt}^{\rm off}$), and if one also sets $\kappa=0$, then $ M_{\rm unit}$ describes a unit cell of the ideal atomic mirror configuration. This yields the Lorentzian reflection\,(\ref{refl1}). To recover the ideal situation, our goal is then to choose $\kappa$ such that it cancels the presence of $\kappa_\mu$ to lowest order. It can readily be shown that
\begin{equation}\label{form_kappa} \kappa=\kappa_e+\kappa_s\quad \mathrm{with} \quad \kappa_\mu=\frac{\Gamma_{\rm 1D}^\mu}{\pi\,\Delta_\mu}\,.\end{equation}
That is, both the off-resonant sublattice $e$- and $s$-transition can contribute to the formation of reflectance dips. Besides the adjustment of the inter-atomic propagation phase by $\kappa$, reducing its magnitude e.g. by increasing the detunings $\Delta_\mu$ can be utilized to suppress the dip formation. As shown in Fig.\,\ref{b_scaling}\,(a), an adjustment of the inter-atomic phase by $\kappa$, which eliminates the dips in the reflectance resonances, prevents the reflectance of a probe pulse from saturating with an increasing optical depth.

\section{Numerical reflectance and transmittance calculation}\label{sec_numimp}
We here illustrate the procedure used for numerically calculating the reflectance and transmittance of a probe field $\mathcal{E}_{\rm in}(z,t)$. It is based on the spin-model description\,\cite{caneva15}, i.e. a time evolution of the atomic (spin) state $\ket{\psi}$ under an effective non-Hermitian Hamiltonian $\mathcal{H}$, and the subsequent reconstruction of the field based on input-output relations (see Section\,\ref{sec_config} for the effective spin Hamiltonian and the general form of the input-output operators). In order to reduce the numerical complexity, we truncate the atomic wavefunction $\ket{\psi}$ to a maximum of two excitations in $e$ or $s$, i.e. we keep only the ground state $\ket{g_0}=\ket{g}^{\otimes N}$ with all atoms in $g$, the single-excitation states \{$\ket{e_m}$, $\ket{s_m}$\} with one atom (e.g. atom $m$) excited, and the two-excitation states \{$\ket{e_m e_n}$, $\ket{s_m s_n}$, $\ket{e_m s_n}$\} with two excited atoms  (e.g. $m$ and $n$). For that approximation to be valid, i.e. for higher order excitations to be negligible, we restrict the probe field $\mathcal{E}_{\rm in}(z,t)$ to weak coherent fields of photon number $\bar{n}\ll1$.
The wavefunction is initialized in the ground state $\ket{\psi(0)}=\ket{g_0}$ or in a single `gate' excitation state $\ket{\psi(0)}=\ket{s'}$ ($s$-branch) or $\ket{\psi(0)}=\ket{e'}$ ($e$-branch). These latter states can be superpositions of single excitation states, e.g. the subradiant state configuration introduced in section\,\ref{sec_condphoton}, or non-decaying ancilla states as used for the reflectance spectra calculations.  Subsequently the time evolved atomic state $\ket{\psi(t)}$ is numerically calculated based on the Schr\"odinger equation $(\mathrm{d}/\mathrm{d}t)\ket{\psi(t)}=-i\,\mathcal{H}\,\ket{\psi(t)}$.

The reflected $I_r$ and transmitted $I_t$ field intensities at a specific time $t$ and position $z$  then follow from the input-output relations. For example, the transmitted field $\mathcal{E}_+(z_e,t)=\mathcal{E}_{\rm in}(z_e,t)\,\mathbbm{1}+i\,(\Gamma_{\rm 1D}/2)\,\sum_{m=1}^N\sigma_{ge}^m\,e^{i\,k\,(z_e-z_m)}$, where $z_e=N\,a$ denotes the end position of the atomic chain. Thus, the transmitted intensity follows as
\begin{equation}\label{fint}  I_t(z_e,t)=\ex{\psi(t)|\,\mathcal{E}_+(z_e,t)^\dagger\,\mathcal{E}_+(z_e,t)\,|\psi(t)} \end{equation}
and involves the calculation of two-body atomic correlations of the form $\ex{\psi(t)|\,\sigma_{eg}^m\sigma_{ge}^n\,|\psi(t)}$.  The reflected field intensity follows analogously by simply replacing the field operators with the ones of inverse propagation $\mathcal{E}_-(z=0,t)=i\,(\Gamma_{\rm 1D}/2)\sum_{m=1}^N\sigma_{ge}^m e^{ikz_m}$. The transmittance $\mathcal{T}$ then follows by normalizing the transmitted intensity by the input intensity. For a cw input field $\mathcal{E}_{\rm in}(z,t)=\mathcal{E}_0\,e^{i\,(kz-\delta\,t)}$, e.g. for the calculation of the transmittance spectra, this is performed by evaluating (\ref{fint}) in the steady state regime (i.e. at a time when $I_t$ reaches a constant value $I_t^{\rm ss}$ in time), and the transmittance then follows as $\mathcal{T}=I_t^{\rm ss}/|\mathcal{E}_0|^2$. For a signal pulse, e.g. $\mathcal{E}_{\rm in}(z,t)=\mathcal{E}_0\,\sin^2(\pi\,t/(2 t_0))\,e^{i\,(k z-\delta t)}$, it is defined by $\mathcal{T}=(\int I_t(z_e,t')\,\mathrm{d}t')/(\int |\mathcal{E}_{\rm in}(t')|^2\,\mathrm{d}t')$, where in addition the background field of the dissipative `gate excitation' is subtracted as discussed below. The reflectances $\mathcal{R}$ and higher order correlations as $g^{(2)}(0)$ are calculated analogously. 

As the time evolution follows from the Schr\"odinger equation in the wavefunction picture, the dissipative nature, i.e. the emission into free space ($\Gamma'$) and into the waveguide ($\Gamma_{\rm 1D}$), is only accounted for by the norm decay of the wavefunction evolving under the non-Hermitian Hamiltonian $\mathcal{H}$. In particular, quantum jumps, e.g. a jump from $\ket{e}$ to $\ket{g}$ by the emission of a photon, are not correctly accounted for. Thus, either the occurrence of these jumps must be infrequent, which implies that the wavefunction norm decay on the timescale of interest is small, or the state after a jump event (e.g. $\ket{g}$) must have a negligible impact on the observables of interest $\mathcal{R}$ and $\mathcal{T}$.

\begin{figure}[htb]
\begin{centering}
\includegraphics[scale=0.3]{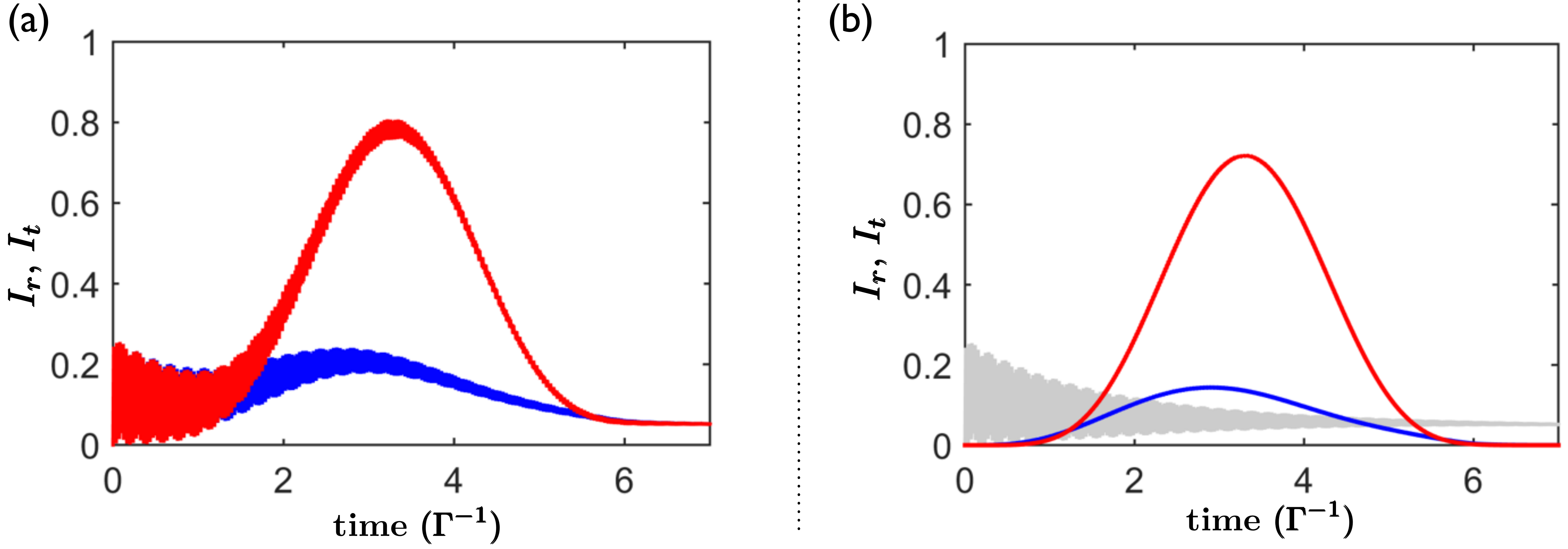}
\caption{\label{b_bill} Reflected $I_r$ (red) and transmitted $I_t$ (blue) signal pulse intensity in time for a signal pulse  $\tilde{\mathcal{E}}_{\rm in}(t)=\mathcal{E}_0\,\sin^2(\pi t/(2 t_0))$  [$0\leq t\leq 2\,t_0$] centered around the broader $s$-resonance of Fig.\,\ref{b_decspec}\,(b) and for an initial `gate' excitation in $s$ prepared at $t=0$. The intensities are normalized by $|\mathcal{E}_0|^2$. (a) Including the background field leaked by the initial excitation. (b) With subtracted background field as explained in the main text. Gray lines represent the background field intensity alone.   \textit{Parameters: } $N=200$, $\Gamma_{\rm 1D}=0.5\,\Gamma'$, $\delta_c=470\,\Gamma'$, $\Omega=94\,\Gamma'$, $\mathcal{J}=235\,\Gamma'$, $\delta=514.9\,\Gamma'$, $t_0=2.95\,\Gamma'^{-1}$, $\mathcal{E}_0=2\,${e-4}\,$\Gamma'$ and  $\kappa=8\,${e-4}.  } 
\end{centering}
\end{figure}

For a weak coherent field input and in the absence of a `gate' excitation, the total excited state population $\wp_{\rm ex}$ remains small and the occurrence of quantum jumps (e.g. with rate $\sim\wp_{\rm ex}\,\Gamma'$) can be neglected on the considered timescale (e.g. the time to reach a steady state that forms on a timescale $\sim \Gamma'^{-1}$).  The case of an initial `gate' excitation is more subtle as it involves a full excitation, whose decay is generally non-negligible.
In our specific case, where our observable of interest is reflection of a signal field, including jumps would yield a negligible change. In particular, a jump to $\ket{g_0}$ causes the system to be almost perfectly transmitting. This originates from the fact that the resulting state following the jump corresponds to a configuration far-detuned from the probe field, which is assumed to be tuned to the two-excitation reflectance resonance. Therefore the field then does not interact with the atomic lattice any more and thus does not contribute to the reflectance signal. The decay of reflectance associated with the gate excitation decay is correctly accounted for by the wavefunction norm decay and does not require one to consider jump events.

Despite the `gate' excitation being of subradiant nature with respect to the waveguide emission, the field associated with its decay, named `background field' in the following, can still be significant compared to the field of a weak-coherent `signal' pulse. In contrast, for a true single photon pulse it would be negligible in magnitude. As all of our calculations are performed with weak coherent input states for the signal, we must subtract out the part of the intensity that arises from leakage of the gate into the waveguide.
The background subtracted pulse reflectance can be obtained in the following way: Calculating the output field intensity in the absence of a `signal' pulse, i.e. for $\mathcal{E}_{\rm in}\equiv0$ and an initial gate excitation e.g. $\ket{\psi(0)}=\ket{s'}$, and subtracting that intensity from the combined intensity in its presence ($\mathcal{E}_{\rm in}\neq0$,  $\ket{\psi(0)}=\ket{s'}$).
The method has been found successful in recovering the expected pulse shapes and scalings even for background fields that are orders of magnitude larger than the reflected `signal' field. As an example, we illustrate in Fig.\,\ref{b_bill} the reflected and transmitted intensity of a `signal' pulse in the presence of a decaying `gate' excitation in $s$. In Fig.\,\ref{b_bill}\,(a) the total combined intensity of signal pulse field and the field leaked by the gate excitation is shown. The signal pulse shape can be identified on top of a fast-oscillating background, where the oscillations are due to the non-eigenstate nature of the initial excitation. The background subtracted intensities are depicted in Fig.\,\ref{b_bill}\,(b) along with the background field alone. A clear separation of these individual components is obtained; both the oscillating nature is removed for the background subtracted pulse intensities and a proper normalization is observed, i.e. the intensity is (approaches) zero at the beginning and end of the signal pulse.

\section*{References}
\bibliography{mirror_ref}

\end{document}